\documentclass[10pt]{article}

\usepackage{graphics}
\usepackage{amsmath}
\usepackage{amsfonts,amsbsy,amssymb}

\def\empile#1\above#2{\mathrel{\mathop{\kern 0pt#1}\limits_{#2}}}

\newcommand{\sll}{\raise.15ex\hbox{$/$}\kern-.43em\hbox{$l$}}
\newcommand{\slepsilon}{\raise.15ex\hbox{$/$}\kern-.53em\hbox{$\epsilon$}}
\newcommand{\slvarepsilon}{\raise.15ex\hbox{$/$}\kern-.53em\hbox{$\varepsilon$}}
\newcommand{\slL}{\raise.15ex\hbox{$/$}\kern-.53em\hbox{$L$}}
\newcommand{\slP}{\raise.15ex\hbox{$/$}\kern-.53em\hbox{$P$}}
\newcommand{\slp}{\raise.1ex\hbox{$/$}\kern-.63em\hbox{$p$}}
\newcommand{\slq}{\raise.1ex\hbox{$/$}\kern-.53em\hbox{$q$}}
\newcommand{\slv}{\raise.1ex\hbox{$/$}\kern-.63em\hbox{$v$}}
\newcommand{\slR}{\raise.15ex\hbox{$/$}\kern-.53em\hbox{$R$}}
\newcommand{\slQ}{\raise.15ex\hbox{$/$}\kern-.53em\hbox{$Q$}}
\newcommand{\slK}{\raise.15ex\hbox{$/$}\kern-.53em\hbox{$K$}}
\newcommand{\slk}{\raise.15ex\hbox{$/$}\kern-.53em\hbox{$k$}}
\newcommand{\slSigma}{\raise.15ex\hbox{$/$}\kern-.53em\hbox{$\Sigma$}}
\newcommand{\slcalP}{\raise.15ex\hbox{$/$}\kern-.63em\hbox{$\cal P$}}
\newcommand{\slA}{\raise.15ex\hbox{$/$}\kern-.73em\hbox{$A$}}
\newcommand{\slbfA}{\raise.15ex\hbox{$/$}\kern-.73em\hbox{${\imb A}$}}
\newcommand{\slpartial}{\raise.15ex\hbox{$/$}\kern-.53em\hbox{$\partial$}}
\newcommand{\sla}{\raise.15ex\hbox{$/$}\kern-.53em\hbox{$a$}}
\newcommand{\slb}{\raise.15ex\hbox{$/$}\kern-.53em\hbox{$b$}}
\newcommand{\slc}{\raise.15ex\hbox{$/$}\kern-.53em\hbox{$c$}}
\newcommand{\slD}{\raise.15ex\hbox{$/$}\kern-.53em\hbox{$D$}}
\newcommand{\slC}{\raise.15ex\hbox{$/$}\kern-.53em\hbox{$C$}}

\def\p{{\boldsymbol p}}
\def\q{{\boldsymbol q}}
\def\l{{\boldsymbol l}}
\def\k{{\boldsymbol k}}

\def\x{{\boldsymbol x}}
\def\y{{\boldsymbol y}}

\def\r{{\boldsymbol r}}
\def\z{{\boldsymbol z}}
\def\b{{\boldsymbol b}}

\def\u{{\boldsymbol u}}
\def\v{{\boldsymbol v}}
\def\P{{\boldsymbol P}}

\def\wt{\widetilde}
\def\bs{\boldsymbol}

\begin{document}

\thispagestyle{empty}
\title {\bf Quark pair production in high\\
energy pA collisions: General features}

\author{Hirotsugu Fujii$^{(1)}$, Fran\c cois Gelis$^{(2)}$, Raju Venugopalan$^{(3)}$}
\maketitle
\begin{center}
\begin{enumerate}
\item Institute of Physics\\ 
  University of Tokyo, Komaba \\ 
  Tokyo 153-8902, Japan
\item Service de Physique Th\'eorique\footnote{URA 2306 du CNRS.}\\
  B\^at. 774, CEA/DSM/Saclay\\
  91191, Gif-sur-Yvette Cedex, France
\item Physics Department\\
  Brookhaven National Laboratory\\
  Upton, NY 11973, USA
\end{enumerate}
\end{center}

\begin{abstract}
  \noindent A consistent treatment of both multiple scattering and
  small $x$ quantum evolution effects on pair production in high energy
  pA collisions is feasible in the framework of the Color Glass
  Condensate (CGC)~\cite{BlaizGV2}. We first discuss the properties of
  quark pair production in the classical effective theory where only
  multiple scattering effects are included. Explicit results are given
  for pair production as a function of the invariant mass of pairs,
  the pair momenta, the atomic mass number $A$ and the quark mass.  We
  relate the logarithms that appear in our formulation of pair
  production to logarithms that appear in the limit of collinear
  factorization in QCD. Violations of $k_\perp$ factorization and
  medium modifications, as represented by the Cronin effect, are also
  investigated. We next consider how small $x$ quantum evolution
  (shadowing) effects modify the results for pair production. In
  particular, we provide results for the rapidity distribution of
  pairs and the dependence of the Cronin effect on rapidity.  We
  discuss the dependence of our results on the initial conditions for
  small $x$ evolution and comment on its implications for pair
  production at RHIC and the LHC.
\end{abstract}

\section{Introduction}
Proton-nucleus collisions provide a laboratory for studying the
interaction of colored partons with an extended colored medium. Due to
the combined effects of quantum mechanical coherence over extended
longitudinal distances, and the QCD evolution of nuclear
wave-functions, the parton from the projectile can probe both the
density of color charges in the medium, and the multi-parton
correlations that are intrinsic to high parton density components of
the nuclear wave function.  At high energies, the typical momentum
transfer from partons in the medium to the probe is no longer soft and
is characterized by a semi-hard scale\footnote{Throughout this paper,
we simply denote $Q_s$ the saturation momentum of the nucleus. In the
rare instances in which we need to introduce the proton saturation
momentum, we will denote the latter $Q_{s,p}$ in order to distinguish
from that of the nucleus.} $Q_{s}^2 \gg \Lambda_{\rm QCD}^2$.  This
scale, termed the saturation scale, is proportional to the density of
partons in the transverse radius of the nucleus, and grows with
energy~\cite{GriboLR1,MuellQ1,BlaizM1}. Because the running of the
coupling is controlled by this scale, asymptotic freedom tells us that
the coupling of the colored partonic probe should be weak and will
become weaker at higher energies. Therefore, with some effort, one can
hope to learn about the properties of the medium -- in the sense that
one can compute reliably medium effects on the final state observables
that are measured by experiments.

The study of high parton density effects in QCD can be formulated, in
weak coupling, as an effective field theory -- the Color Glass
Condensate
(CGC)~\cite{McLerV1,McLerV2,McLerV3,IancuLM1,IancuLM2,FerreILM1,IancuLM3,IancuV1,Muell4}. The
CGC has been widely applied to study gluon and quark production in
proton-nucleus collisions-for a review, see ref.~\cite{JalilK1}.  An
attractive feature of the CGC effective theory is that one can
quantify what one means by dilute or dense scatterers as a function of
energy and mass number~\cite{BlaizGV1}.  What we mean by
proton-nucleus collisions specifically, is a systematic expansion of
amplitudes to lowest order in the ratio of the saturation momentum of
the proton to the typical transverse momentum exchanged by the proton
in the reaction ($Q_{s,p}/k_{\perp,p} \ll 1$) and all orders in
the ratio of the saturation momentum of the nucleus relative to the
momentum exchanged by the nucleus ($Q_s/k_{\perp,A}$).  At very high
energies, or in very forward kinematics (in the fragmentation region
of the nucleus), the proton saturation scale can be large. In these
kinematics, for fixed impact parameter, proton-nucleus collisions will
be indistinguishable from nucleus-nucleus collisions.

For gluon production in proton-nucleus collisions, the cross-section
can be expressed in $k_\perp$-factorized form as a product of
unintegrated {\it $k_\perp$-dependent} distributions for the proton
and the nucleus, convolved with the hard scattering matrix
element~\cite{KovchM3,Braun1,Braun3,KopelR1,KharzKT1,BlaizGV1}. For a
dilute projectile, such as a proton, the corresponding unintegrated
gluon distribution at lowest order is a leading twist quantity which,
integrated over $k_\perp$, gives the usual leading log collinear gluon
distribution. In contrast, the unintegrated gluon distribution of the
dense target, the nucleus, contains all twist contributions and has no
analogue in the leading twist collinear factorization formalism. In
addition, this remarkable factorization is unlikely to hold beyond
leading order in the expansion in powers of
$Q_{s,p}(x_p)/k_{\perp,p}$~\cite{Balit2,KrasnNV1,KrasnNV2}.

It was shown in ref.~\cite{BlaizGV2} that $k_\perp$-factorization is
broken explicitly in pair production at leading order in
proton-nucleus collisions. Novel 3-point and 4-point
multi-parton (``all twist") correlation functions appear in the
expression for the pair production cross-section. At large transverse
momenta, $k_\perp \gg Q_s$, these expressions simplify~\cite{GelisV1}
and the cross-section smoothly reduces to the $k_\perp$-factorization
result of Collins and Ellis~\cite{ColliE1} and Catani, Ciafaloni and
Hautmann~\cite{CatanCH1}.  The $k_\perp$-factorization formalism of these
authors has been used in several phenomenological studies of heavy
quark production at collider energies~\cite{HagleKSST2,LipatSZ1,YuanC1}. The
magnitude of the breaking of $k_\perp$-factorization for single quark
production was quantified in ref.~\cite{FujiiGV1}.

In this paper, we shall discuss in detail qualitative features of pair
 production that follow from the formalism developed
in ref.~\cite{BlaizGV2}. A full treatment of multiple scattering and
quantum evolution effects at high energies requires a computation of
the previously mentioned 2-point, 3-point and 4-point correlation
functions as a function of $x$, or of the rapidity $Y$ (= $\ln(1/x)$).
These can be determined in full generality (including all leading
logarithms in $x$) by solving the Balitsky-JIMWLK equations for the
small $x$ evolution of multi-parton
correlators~\cite{JalilKLW1,JalilKLW2,JalilKLW3,JalilKLW4,KovneM1,
KovneMW3,Balit1,JalilKMW1,IancuLM1,IancuLM2,FerreILM1,IancuLM3,IancuV1,Muell4}.
This would require an extensive numerical effort -- only preliminary
studies have been performed in this
direction~\cite{RummuW1}. Nevertheless, a considerable deal can be
learnt in certain limits. At large $N$, and for large nuclei, the
small $x$ evolution of the 2-point correlators has a closed form
expression called the Balitsky-Kovchegov (BK)
equation~\cite{Balit1,Kovch1,Kovch3}.  Numerical solutions of the
Balitsky-Kovchegov equation are known in the fixed
coupling~\cite{AlbacAKSW1,Braun4,Lubli1,LevinL1,GolecMS1,GotsmKLMN1}
and running coupling~\cite{RummuW1,AlbacAMSW1}
cases\footnote{\label{foot:noteBK}There are in addition several
analytical studies which capture the key features of the BK equation
\cite{LevinT1,Trian1,MunieP1,MunieP2,MunieP3}.}.  In the large $N$
and large $A$ limit, the 3-point and 4-point correlators also simplify
\cite{KovneW1,Fujii1,FujiiM1,BlaizGV2} and their evolution in $x$ can
be expressed in terms of the solution of the BK equation. These are
computed numerically in this paper. We will also study heavy quark
production in the McLerran-Venugopalan (MV)
model~\cite{McLerV1,McLerV2,McLerV3}, where high parton density
effects contributing to multiple scattering are included but those
arising from small $x$ evolution are not included.  We are thus able
to study quantitatively, in these MV and BK ``mean field" limits, the
interplay of multiple scattering and small $x$ quantum evolution on
pair production.
 
The high parton density effects we discuss here include contributions
categorized as ``higher twist" in the established language of
collinear factorization. These effects are difficult to treat in that
framework and therefore a matching of the two formalisms is difficult.
However, one can match the two formalisms at large transverse momenta
(specifically, when $Q_s^2 \ll p_\perp^2, M^2$) where
$k_\perp$-factorization formalism is recovered. We shall discuss in
general how the logarithms that appear in the $k_\perp$-factorization
framework can be related to
DGLAP~\cite{Doksh1,AltarP1,GriboL1,GriboL2} logarithms that arise in
collinear factorization. Similar considerations arose in other
specific processes~\cite{DumitM1,DumitJ1,GelisJ1,GelisJ2}.

The formalism discussed here is valid when at least one of the sources
is dilute, as in pA collisions. When the dilute source becomes dense
(either by increasing the energy of the proton or in a nucleus-nucleus
collision), higher order corrections in $Q_{s,p}/k_\perp$ contribute
significantly. Thus, in particular for nucleus-nucleus collisions,
pair production has to be computed numerically; first results were
obtained recently~\cite{GelisKL1,GelisKL2}.

We will not address phenomenological applications of our approach to
pair production in pA collisions in this paper.  There are interesting
results from D-Au collisions at RHIC~\cite{AdlerA1,AdamsA1} and in pA
collisions at the SPS~\cite{AlessA1,AlessA2,WohriA1}.  Some of this data
has been studied previously in models based on the CGC
approach~\cite{KharzT3,KharzT4}. An interesting comparative
study of several models of $J/\psi$ production in heavy ion collisions
and relevant references can be found in ~\cite{DuraeNN1}. We will
address the existing data and make predictions for future data on pair
production at RHIC and LHC in a follow up to this
work~\cite{FujiiGV2}.

This paper is organized as follows. In section \ref{sec:cs}, we state
the key results obtained in our previous derivation~\cite{BlaizGV2} of
the pair production and single quark cross-sections. These were
derived as a function of the momenta and rapidity of the quark and the
anti-quark and we will re-express these in terms of the pair momenta,
invariant mass and rapidity. In section \ref{sec:corr}, we shall
discuss the properties of the multi-parton correlators and present
results for their evolution as a function of energy in the large $N$
and large $A$ limit. In section 4, we will present results for the
pair cross-sections. The results in this section are presented, in the
CGC framework, for a) the McLerran-Venugopalan model and b) for the
Balitsky-Kovchegov mean field model of the CGC. As mentioned
previously, the former is a reasonable model in the kinematic region
where multiple scattering effects are important and small $x$ quantum
evolution can be neglected.  We discuss, in this model, pair
distributions as a function of the pair momenta, rapidity and
invariant mass.  In particular, we show that the behavior of the
spectra at large momenta can be understood analytically. The
logarithms that appear in these expressions have the same origin as
the collinear logs that appear in collinear factorization formalism of
perturbative QCD. At smaller momenta, $k_\perp$ factorization is
broken explicitly. We demonstrate the dependence of this breaking on
the invariant mass and momenta of the pair.  In the MV model, the
saturation scale $Q_s$ depends on the size of the nucleus. We study
the dependence of the pair cross-sections on $Q_s$ and on the quark
mass.  We next discuss the behavior of these quantities when small x
quantum evolution is turned on. These shadowing effects are obtained
by computing the rapidity dependence of correlators with the
Balitsky-Kovchegov equation.  We compute the rapidity distribution and
the invariant mass and momentum distribution of pairs. In particular,
we study the Cronin effect for quark pairs and its variation as a
function of rapidity. We comment on the dependence of the small x
quantum evolution on the large $x$ initial conditions. Section 6
summarizes our conclusions.

\section{Quark cross-sections}
\label{sec:cs}
\subsection{Generalities}
\begin{figure}[htbp]
\begin{center}
\resizebox*{!}{5cm}{\includegraphics{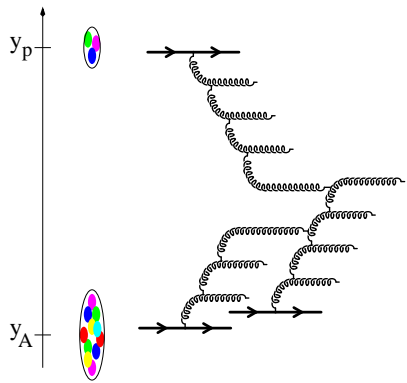}}
\end{center}
\caption{\label{fig:collision}Cartoon representing gluon production in
proton-nucleus collisions in the Color Glass Condensate framework.}
\end{figure}

\begin{figure}[htbp]
\begin{center}
\resizebox*{!}{2cm}{\includegraphics{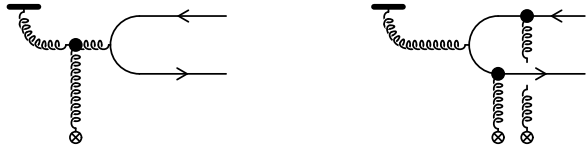}}
\end{center}
\caption{\label{fig:qqbar-prod} The two types of terms
that enter in the $Q\overline{Q}$ production amplitude in pA
collisions. The gluon emitted by the proton can either emit the pair
after the collision with the nucleus (left diagram) or before
colliding with the nucleus (diagram on the right). The black blob
denotes multiple scatterings resummed via Wilson lines.}
\end{figure}

In the CGC formalism, the proton-nucleus collision is described as a
collision of two classical fields originating from color sources
representing the large $x$ degrees of freedom in the proton and the
nucleus. A schematic representation of the collision in this formalism
is shown in fig.~\ref{fig:collision}.
As suggested by the figure, the color source distribution generating
the classical field in each projectile has been evolved from some
initial valence distribution at large $x$, in the leading logarithmic
approximation in $x$, to the rapidity of interest in the collision. The
gauge fields of gluons produced in the collision are determined by
solving the Yang-Mills equations
\begin{equation}
[D_\mu,F^{\mu\nu}] = J^\nu \, .
\label{eq:YM}
\end{equation}
Here $J^\nu$ is the color current of the sources, which can expressed
at leading order in the sources as
\begin{equation}
J_a^\nu = g\delta^{\nu+}\delta(x^-)\,\rho_{p,a}(\x_\perp) + g\delta^{\nu-}\delta(x^+)\rho_{_A,a}(\x_\perp) \, ,
\label{eq:current}
\end{equation}
where $\rho_p$ is the number density of ``valence" partons in the
proton moving in the $+z$ direction at the speed of light.  Likewise,
$\rho_{_A}$ is the number density of ``valence" partons in the nucleus
moving in the opposite light cone direction. The previous two
equations must be supplemented by a gauge fixing condition, and by the
covariant conservation of the current~:
\begin{equation}
[D_\nu,J^\nu]=0\; .
\label{eq:cur-conv}
\end{equation}
The latter equation in general implies that eq.~(\ref{eq:current}) for
the current receives corrections that are of higher order in the
sources $\rho_p$ and $\rho_{_A}$, because of the radiated field. The
solution of eqs.~(\ref{eq:YM}), (\ref{eq:current}) and
(\ref{eq:cur-conv}) has been determined to all orders in both sources
only numerically~\cite{KrasnNV1,KrasnNV2,Lappi1}. To lowest order in
the proton source (as appropriate for a dilute proton source) and to
all orders in the nuclear source, analytical results are available and
an explicit expression for the gauge field to this order, in Lorentz
gauge, is given\footnote{This solution has been derived in
\cite{GelisM1} in the light-cone gauge of the proton, and in
\cite{DumitM1} in the gauge $x^+A^-+x^-A^+=0$.} in
ref.~\cite{BlaizGV1}. The amplitude for pair production to this order
is obtained by computing the quark propagator in the background
corresponding to this gauge field~\cite{BlaizGV2}. The diagrams
corresponding to these insertions are shown in fig.~\ref{fig:qqbar-prod}.

The probability for producing a single\footnote{The probability of
producing two pairs or more is parametrically of higher order in the
proton source. Since we assume that $\rho_p$ is small, producing a
single pair is thus the dominant process.} $q\bar{q}$ pair in the
collision can be expressed as
\begin{equation}
P_1[\rho_p,\rho_{_A}]=
\int\frac{d^3\q}{(2\pi)^3 2E_\q}
\int\frac{d^3\p}{(2\pi)^3 2E_\p}
\left|{\cal M}_{_{F}}(\q,\p)\right|^2\; ,
\label{eq:P1-def}
\end{equation}
where ${\cal M}_{_{F}}(\q,\p)$ is the amputated time-ordered quark
propagator in the presence of the classical field.  The argument
$[\rho_p,\rho_{_A}]$ indicates that this is the production probability
in one particular configuration of the color sources. In order to turn
this probability into a cross-section, one must first average over the
initial classical sources $\rho_p$ and $\rho_{_{A}}$ respectively with
the weights~\footnote{These weight functionals are normalized to
ensure that their respective path integrals over the sources are
unity.} $W_p[x_p,\rho_p], W_{_{A}}[x_{_A},\rho_{_{A}}]$ and
subsequently integrate over all the impact parameters $\b$, to obtain,
\begin{equation}
\sigma=\int d^2\b \int[D\rho_p][D\rho_{_{A}}]
\,W_p[x_p,\rho_p]\,W_{_{A}}[x_{_A},\rho_{_{A}}]
P_1[\rho_p,\rho_{_A}]\; .
\label{eq:source-avg}
\end{equation}
This formula incorporates both multiple scattering effects and quantum
effects. The multiple scattering effects are in part included in the
classical field (the solution of Yang-Mills equations), in part in the
propagator of the quark in this classical field, and also in the
evolution with $x$ of the source distributions.  The quantum effects
are included in the evolution of the weight functionals, $W_p$ and
$W_{_{A}}$, of the target and projectile with $x$.  The arguments
$x_p$ and $x_{_A}$ denote the scale in $x$ separating the large-$x$
static sources from the small-$x$ dynamical fields. In the
McLerran-Venugopalan model, the functional $W_{_{A}}$ that describes
the distribution of color sources in the nucleus is a Gaussian in the
color charge density~\footnote{This is true modulo a term proportional
to the Cubic Casimir, that is parametrically suppressed relative to
the Gaussian term at large $A$~\cite{JeonV1,JeonV2}--see eq.~\ref{eq:GW}.} in $\rho_{_{A}}$
\cite{McLerV1,McLerV2,McLerV3}.  Having a Gaussian distribution of
sources, in our framework, is equivalent to the Glauber model of
independent multiple scattering~\cite{BlaizGV1}.  In general, this
Gaussian is best interpreted as the initial condition for a
non-trivial evolution of $W_{_{A}}[x_{_A},\rho_{_{A}}]$ with
$x_{_A}$. This evolution is described by a Wilson renormalization
group equation -- the JIMWLK
equation~\cite{JalilKLW1,JalilKLW2,JalilKLW3,JalilKLW4,KovneM1,
KovneMW3,Balit1,JalilKMW1,IancuLM1,IancuLM2,FerreILM1}. We will
discuss evolution equations further in the following section.

\subsection{Differential pair cross-section}
The pair production cross-section, for quarks (anti-quarks) of momenta
$\q_\perp$ ($\p_\perp$) and rapidities $y_q = 1/2 \ln(q^+/q^-)$ ($y_p
= 1/2\ln(p^+/p^-)$), derived in \cite{BlaizGV2} can be expressed as
\begin{eqnarray}
&&\frac{d\sigma}{d^2\p_\perp d^2\q_\perp dy_p dy_q}=
\frac{\alpha_s^2 N}{8\pi^4 (N^2-1)}
\int\limits_{\k_{1\perp},\k_{2\perp}}\!\!\!
\frac{\delta(\p_\perp+\q_\perp-\k_{1\perp}-\k_{2\perp})}
{\k_{1\perp}^2 \k_{2\perp}^2}\,\varphi_p(\k_{1\perp})
\nonumber\\
&&\qquad\qquad\!\!\!\!\!
\times\Bigg\{
\int\limits_{\k_\perp,\k_\perp^\prime}
\!\!\!\!\!
{\rm tr}_{\rm d}
\Big[(\slq\!+\!m)T_{q\bar{q}}(\slp\!-\!m)
\gamma^0 T_{q\bar{q}}^{\prime\dagger}\gamma^0\Big]
\phi_{_A}^{q\bar{q},q\bar{q}}
(\k_{2\perp};\k_\perp,\k_\perp^\prime)
\nonumber\\
&&\qquad\qquad\;\;
+\int\limits_{\k_\perp}
\!
{\rm tr}_{\rm d}
\Big[(\slq\!+\!m)T_{q\bar{q}}(\slp\!-\!m)
\gamma^0 T_{g}^{\dagger}\gamma^0 + {\it h.c.}\Big]
\phi_{_A}^{q\bar{q},g}(\k_{2\perp};\k_\perp)
\nonumber\\
&&\qquad\qquad\qquad
+{\rm tr}_{\rm d}
\Big[(\slq\!+\!m)T_{g}(\slp\!-\!m)\gamma^0 T_{g}^{\dagger}\gamma^0\Big]
\phi_{_A}^{g,g}(\k_{2\perp})
\Bigg\}
\; .
\label{eq:cross-section}
\end{eqnarray}
where we denote \footnote{The momenta $p$ and $q$ of the produced
  particles have not been listed among the arguments of these objects
  to ensure the equations are more compact.}
\begin{eqnarray}
&&T_{q\bar{q}}(\k_{1\perp},\k_{\perp})\equiv 
\frac{\gamma^+(\slq-\slk+m)\gamma^-(\slq-\slk-\slk_1+m)\gamma^+}
{2p^+[(\q_\perp\!-\!\k_\perp)^2+m^2]+2q^+[(\q_\perp\!-\!\k_\perp\!-\!\k_{1\perp})^2+m^2]}
\nonumber\\
&&T_{g}(\k_{1\perp})\equiv 
\frac{\slC_{_{L}}(p+q,\k_{1\perp})}{(p+q)^2}
\; .
\label{eq:Tqqbar-Tg}
\end{eqnarray}
Note that $T_{q\bar{q}}^\prime\equiv
T_{q\bar{q}}(\k_{1\perp},\k_{\perp}^\prime)$ and $C_L^\mu$ is the
well-known Lipatov vertex defined as
\begin{equation}
C_{_L}^+(q,\k_{1\perp}) \equiv \frac{-k_{1\perp}^2}{q^-}+ q^+\,\,;\,\,
C_{_L}^-(q,\k_{1\perp}) \equiv \frac{k_{2\perp}^2}{q^+}-q^-\,\,;\,\,
C_{_L}^i(q,\k_{1\perp}) \equiv -2k_1^i +q^i \, ,
\label{eq:Lipatov}
\end{equation}
with $\k_{2\perp} \equiv \q_\perp - \k_{1\perp}$. 

$\varphi_p$ is the unintegrated gluon distribution of the proton and
the various $\phi_{_A}$'s are unintegrated distributions describing
the target. As stated previously, we are working to lowest order in
the color charge density $\rho_p$ of the proton. The proton
unintegrated distribution is given by~\footnote{In the MV model, the
correlator of {\it number densities} is
$\left<\rho_p^a(0)\rho_p^a(\x_\perp)\right>= \mu_A^2
(N^2-1)\delta^{(2)}(\x_\perp-\y_\perp)$, where $\mu_A^2 =
\frac{A}{2\pi R_A^2}$.  Often, in the literature, the correlator of
{\it charge densities} ${\tilde \mu}_A^2$ is used: ${\tilde \mu}_A^2 =
g^2 \mu_A^2$. The unintegrated gluon distribution in
eq.~(\ref{eq:corr-proton}) is normalized such that the leading log gluon
distribution in the proton satisfies $$x G_p (x,Q^2) = {1\over
4\pi^3}\, \int_0^{Q^2} d l_\perp^2 \varphi_p(l_\perp)\; .$$}
\begin{equation}
\varphi_p(\l_\perp)\equiv \frac{\pi^2 R_p^2\,g^2}{l_\perp^2}
\int_{\x_\perp}
e^{i\l_\perp\cdot\x_\perp}\left<\rho_p^a(0)\rho_p^a(\x_\perp)\right>\; .
\label{eq:corr-proton}
\end{equation}
The corresponding nuclear ``unintegrated distributions'' are defined
to be
\begin{eqnarray}
&&\phi_{_{A}}^{g,g}(\l_{\perp})\equiv \frac{\pi^2 R_{_A}^2 l_\perp^2}{g^2 N}
\int_{\x_\perp}
e^{i\l_\perp\cdot\x_\perp}
\,{\rm tr}\,\left<U(0)U^\dagger(\x_\perp)\right>\; ,
\nonumber\\
&&\phi_{_A}^{q\bar{q},g}(\l_\perp;\k_\perp)\equiv
\frac{2\pi^2 R_{_A}^2 l_\perp^2}{g^2 N}
\int_{\x_\perp,\y_\perp}\!\!\!\!\!\!\!\!\!\!
e^{i(\k_\perp\cdot\x_\perp+(\l_\perp-\k_\perp)\cdot\y_\perp)}
\nonumber\\
&&\qquad\qquad\qquad\times
{\rm tr}
\left<
{\wt U}(\x_\perp)t^a {\wt U}^\dagger(\y_\perp) t^b U_{ba}(0)
\right>\nonumber\\
&&\phi_{_A}^{q\bar{q},q\bar{q}}(\l_\perp;\k_\perp,\k_\perp^\prime)\equiv \frac{2\pi^2 R_{_A}^2 l_\perp^2}{g^2 N}
\int_{\x_\perp^\prime,\y_\perp,\y_\perp^\prime}\!\!\!\!\!\!\!\!\!\! e^{i(-\k_\perp\cdot\y_\perp+\l_\perp\cdot (\y_\perp-\y_\perp^\prime)+\k_\perp^\prime\cdot(\y_\perp^\prime-\x_\perp^\prime))}
\nonumber\\
&&\qquad\qquad\qquad\times
{\rm tr}
\left<{\wt U}(0)t^a {\wt U}^\dagger(\y_\perp){\wt U}(\x_\perp^\prime)t^a {\wt U}^\dagger(\y_\perp^\prime)
\right> \, .
\label{eq:corr-nucleus}
\end{eqnarray}

The matrices $U$ are path ordered exponentials along the light cone
longitudinal extent corresponding to partonic configurations of the
nucleus in the infinite momentum frame:
\begin{eqnarray}
U(\x_\perp)\equiv {\cal P}_+ \exp\left[-ig^2\int_{-\infty}^{+\infty}
dz^+ \frac{1}{{\bs\nabla}_{\perp}^2}\,\rho_{_A}(z^+,\x_\perp)\cdot T
\right] \, ,
\end{eqnarray}
where the $T^a$ are the generators of the adjoint representation of
$SU(N)$ and ${\cal P}_+$ denotes a ``time ordering'' along the $z^+$
axis. The ${\wt U}$ have an identical definition, with the generators
$T^a$ replaced by the generators $t^a$ in the fundamental
representation of $SU(N)$. We remind the reader that the expectation
values $\big<\cdots\big>$ here correspond to the averages over the
sources $\rho_p$ in the case of $\varphi_p$ and $\rho_{_A}$ in the case
of the $\phi_{_A}$'s.

In the definition of these distribution functions, the momentum
$\l_\perp$ is the total transverse momentum exchanged between the
nucleus and the probe (either a gluon or the $Q\overline{Q}$ state),
and $\k_\perp,\k^\prime_\perp$ are the momenta exchanged between the
quark and the nucleus. These conventions can be visualized in the
diagrams \setbox1\hbox to
3cm{\hfil\resizebox*{3cm}{!}{\includegraphics{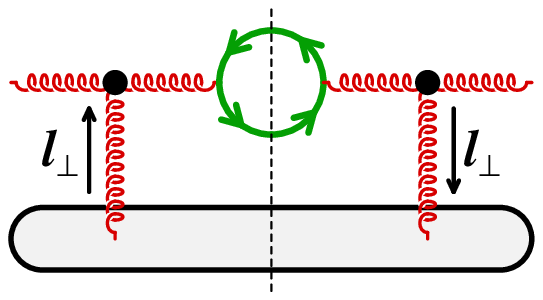}}\hfil}
\setbox2\hbox to
3cm{\hfil\resizebox*{3cm}{!}{\includegraphics{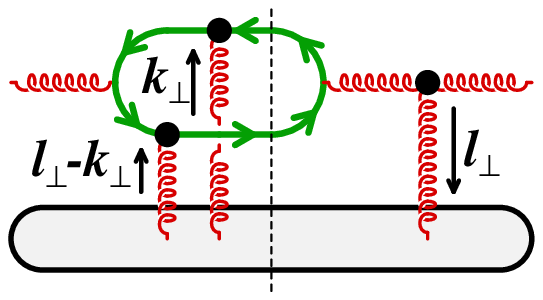}}\hfil}
\setbox3\hbox to
3cm{\hfil\resizebox*{3cm}{!}{\includegraphics{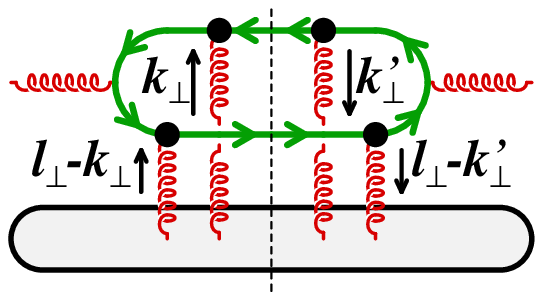}}\hfil}
\begin{eqnarray}
\phi_{_{A}}^{g,g}(\l_{\perp})
&=&\;\;
\raise -5mm\box1\nonumber\\
&&\nonumber\\
\phi_{_A}^{q\bar{q},g}(\l_\perp;\k_\perp)
&=&\;\;
\raise -5mm\box2\nonumber\\
&&\nonumber\\
\phi_{_A}^{q\bar{q},q\bar{q}}(\l_\perp;\k_\perp,\k_\perp^\prime)
&=&\;\;
\raise -5mm\box3
\end{eqnarray}

It is transparent that eq.~(\ref{eq:cross-section}) for the pair
production cross-section is not in general $k_\perp$-factorizable into
a simple convolution of unintegrated parton distributions from the
proton and the nucleus. While one can still factorize out the proton
unintegrated distribution, the nucleus is now represented by the
distributions $\phi_{_A}^{g,g}$, $\phi_{_A}^{q\bar{q},g}$, and
$\phi_{_A}^{q\bar{q},q\bar{q}}$, which are respectively 2-point,
3-point and 4-point correlators of Wilson lines in the color field of
the nucleus. These multi-parton correlation functions contain all
twists and are in general rather complicated. They however, in all
generality, satisfy the sum rule,
\begin{equation}
\int\limits_{\k_\perp,\k_\perp^\prime}
{\phi_{_A}^{q\bar{q},q\bar{q}}(\l_\perp;\k_\perp,\k_\perp^\prime)}
=
\int\limits_{\k_\perp}
{\phi_{_A}^{q\bar{q},g}(\l_\perp;\k_\perp)}
=
{\phi_{_A}^{g,g}(\l_\perp)}\; .
\label{eq:int-2}
\end{equation}

\subsection{Leading twist limit}
It is illustrative to consider when one recovers
$k_\perp$-factorization. For Gaussian correlations,
\begin{equation}
\langle \rho_{_{A},a}(\x_\perp)\rho_{_A,a^\prime}(\x_\perp^\prime)\rangle
=
\delta_{aa^\prime}\delta(\x_\perp-\x_\perp^\prime)\mu_{_{A}}^2\; ,
\label{eq:Gauss-corr}
\end{equation}
where $\mu_{_A}^2 =  A/2\pi R_A^2 \propto A^{1/3}$. One obtains, in the leading twist approximation (where the path
ordered exponentials are expanded to lowest order in
$\rho_{_A}/k_\perp^2$),
\begin{eqnarray}
&&\phi_{_A}^{g,g}(\k_{2\perp})\approx
\pi R^2 \left[{\pi d_{_{A}} g^2 } \frac{\mu_{_{A}}^2}{\k_{2\perp}^2}\right]\; ,
\nonumber\\
&&
\phi_{_A}^{g,q\bar{q}}(\k_{2\perp};\k_\perp)
\approx
\frac{\pi R^2}{2}
 \left[{\pi d_{_{A}} g^2} \frac{\mu_{_{A}}^2}{\k_{2\perp}^2}\right]
(2\pi)^2\left[\delta(\k_\perp)\!+\!\delta(\k_\perp\!-\!\k_{2\perp})\right]
\nonumber\\
&&
\phi_{_A}^{q\bar{q},q\bar{q}}(\k_{2\perp};\k_\perp,\k_\perp^\prime)\approx
\pi R^2 \left[{\pi d_{_{A}} g^2 } \frac{\mu_{_{A}}^2}{\k_{2\perp}^2}\right]
\nonumber\\
&&\qquad\qquad\times
(2\pi)^4\Big[
\frac{C_{_{F}}}{N} \left(
\delta(\k_\perp-\k_{2\perp})\delta(\k_\perp^\prime-\k_{2\perp})
+
\delta(\k_\perp)\delta(\k_\perp^\prime)
\right)
\nonumber\\
&&\qquad\qquad\qquad
+\frac{1}{2N^2}
\left(
\delta(\k_\perp-\k_{2\perp})\delta(\k_\perp^\prime)
+
\delta(\k_\perp)\delta(\k_\perp^\prime-\k_{2\perp})
\right)
\Big]\; .\nonumber\\
&&
\label{eq:LT}
\end{eqnarray}
We assume here that the nuclei have a large uniform transverse area
$\pi R^2$. The leading twist expressions for
$\phi_{_{A}}^{g,q\bar{q}}$ and $\phi_{_A}^{q\bar{q},q\bar{q}}$ have a
simple interpretation. At this order, the probe (gluon or $q\bar{q}$
pair) interacts with the nucleus by a single gluon exchange. The two
delta functions in $\phi_{_{A}}^{g,q\bar{q}}$ correspond to the gluon
being attached to the anti-quark line ($\delta(\k_\perp)$ -- there is
no momentum flow from the nucleus to the quark line) or to the quark
line ($\delta(\k_{2\perp}-\k_\perp)$ -- all the momentum from the
nucleus flows on the quark line). An identical interpretation holds
for the four terms of $\phi_{_A}^{q\bar{q},q\bar{q}}$. If one
substitutes these leading twist approximations in
eq.~(\ref{eq:cross-section}), one recovers the leading twist
$k_\perp$-factorized cross-section for pair
production~\cite{ColliE1,CatanCH1,GelisV1}.

\subsection{Kinematics}
We have to specify the multi-parton correlators
$\phi_{_A}^{g,g}(\l_\perp)$,
$\phi_{_A}^{q\bar{q},g}(\l_\perp;\k_\perp)$, and
$\phi_{_A}^{q\bar{q},q\bar{q}}(\l_\perp;\k_\perp,\k_\perp^\prime)$ in
order to compute the pair production cross-section. The following
section will be devoted to a discussion of how one does this. Before
proceeding, however, we will first discuss the kinematic variables in terms of
which the pair production cross-section is specified.

It is convenient to discuss the properties of the produced pairs in
terms of the pair invariant mass $M^2$, the pair rapidity $Y$ and the
pair transverse momentum $\P_\perp$. These are defined in terms of the
momenta and rapidities of the quark and the anti-quark as
\begin{eqnarray}
\P_\perp &=& \p_\perp + \q_\perp
\nonumber\\
 Y &=& 
\frac{1}{2}\ln\left(
\frac{p^++q^+}{p^-+q^-}
\right)=
\frac{1}{2}
\ln\left(
\frac{\omega_p\,e^{y_p} + \omega_q\,e^{y_q}}
{\omega_p\,e^{-y_p}+\omega_q\,e^{-y_q}}
\right)\nonumber \\
M^2 &=& \omega_p^2 + \omega_q^2 + 2\,\omega_p\,\omega_q\,\cosh\left(y_p - y_q\right) - \P_\perp^2 \; .
\label{eq:pair-kinematics}
\end{eqnarray}
Here $\omega_q \equiv \sqrt{\q_\perp^2 + m^2}$ and $\omega_p \equiv
\sqrt{\p_\perp^2 + m^2}$ are the transverse masses of the quark and
the anti-quark respectively, and $y_p \equiv \ln(p^+/p^-)/2$ and
$y_q\equiv \ln(q^+/q^-)/2$ are their respective rapidities. We would
like to express the cross-section in eq.~(\ref{eq:cross-section}) in
terms of the pair kinematical parameters. Firstly, one notices 
that the cross-section in eq.~(\ref{eq:cross-section}) is expressed in
terms of six variables while the pair invariants correspond to four
variables. In other words, a given set of kinematical parameters for
the pair corresponds to a 2-dimensional manifold in the phase-space of
the quark and anti-quark. One writes
\begin{equation}
\frac{d\sigma}{dM^2\, d^2 \P_\perp\,dY} =\!\!\!
\int\limits_0^{\sqrt{\frac{M^2}{4}-m^2}} \!\!\!\!\!\!d{\tilde q}
\int\limits_0^{2\pi}d\phi\;
\frac{{\tilde q}\gamma_x}{\omega_p\,\omega_q \,|\sinh\left(y_p - y_q\right)|}
\;
\frac{d\sigma}{d^2\p_\perp dy_p \,d^2 \q_\perp dy_q} \; .
\label{eq:pair-cross-section}
\end{equation}
In the $Q\overline{Q}$ cross-section that appears under the integral
in eq.~(\ref{eq:pair-cross-section}), the momenta $q$ and $p$ of the
quark and anti-quark are given by
\begin{eqnarray}
&&
q^\mu=
{\bs L_z}(\beta_z)^\mu_\nu \; {\bs L_x}(\beta_x)^\nu_\sigma\; q^\sigma_{\rm cm}
\nonumber\\
&&
p^\mu=
{\bs L_z}(\beta_z)^\mu_\nu \; {\bs L_x}(\beta_x)^\nu_\sigma\; p^\sigma_{\rm cm}
\; ,
\label{eq:trans}
\end{eqnarray}
where $q^\sigma_{\rm cm}$ and $p^\sigma_{\rm cm}$ are the momenta of
the quark and anti-quark in the rest frame of the pair\footnote{We are
here choosing $q^z_{\rm cm}$ to be positive and $p^z_{\rm cm}$ to be
negative. However, the opposite choice is of course allowed as
well. We don't need to consider it explicitly, and multiplying the
final result by a factor $2$ will be sufficient.},
\begin{eqnarray}
&&
q^\sigma_{\rm cm}
=
\left(\frac{M}{2},\tilde{q}\cos\phi,\tilde{q}\sin\phi,
\sqrt{\frac{M^2}{4}-m^2-\tilde{q}^2}\right)
\nonumber\\
&&
p^\sigma_{\rm cm}
=
\left(\frac{M}{2},-\tilde{q}\cos\phi,-\tilde{q}\sin\phi,
-\sqrt{\frac{M^2}{4}-m^2-\tilde{q}^2}\right)\; .
\end{eqnarray}
To proceed from these momenta in the rest frame of the pair to the
corresponding momenta in the laboratory frame in eq.~(\ref{eq:trans}),
two Lorentz boosts must be applied, denoted by ${\bs L_x}$ and ${\bs
L_z}$. Assuming that the pair transverse momentum, $\P_\perp$, is in
the $x$ direction\footnote{This choice is arbitrary, but has no
influence on the result because the cross-section does not depend on
the direction of $\P_\perp$.}, we first apply a boost in the $x$
direction, with a velocity $\beta_x \equiv |\P_\perp| / \sqrt{M^2 +
\P_\perp^2}$ -- hence $\gamma_x\equiv \sqrt{M^2 + \P_\perp^2} / M$. At
this point, the pair has a non-zero transverse momentum, but its $P_z$
(and hence its rapidity) is still zero. A second boost must be applied
in the $z$ direction, in order to bring the pair to the desired
rapidity. The velocity of this boost in the $z$ direction is
$\beta_z\equiv \tanh(Y)$. The other factor in the integrand of
eq.~(\ref{eq:pair-cross-section}) is the Jacobian for the transformation
$\p_\perp,\q_\perp,y_p,y_q \to \P_\perp,Y,M^2,\tilde{q},\phi$.

\section{Multi-parton correlators in the CGC}
\label{sec:corr}
\subsection{Correlators in the MV model}
In the CGC approach, there is a procedure to compute the 2-,\,3- and
4-point multi-parton correlators in eq.~(\ref{eq:corr-nucleus}), in
full generality, in the leading logarithmic approximation in $x$,
starting from a given initial condition at some $x_0$.  However, no
analytic solutions of these evolution equations are available and only
preliminary work has been done in solving them
numerically~\cite{RummuW1}. Fortunately, the multi-parton correlators
simplify greatly in in the large $N$ and large $A$ asymptotic limit of
the theory. While this limit is truly asymptotic, we will assume that
its domain of validity can be extended to finite $A$ and $N$.

In the kinematic domain $\ln(A^{1/3}) \ll y_{_A} \ll A^{1/6}$ where
$y_{_A} \equiv \ln(1/x_{_A})$, the weight functional $W_{_{A}}[x_{_A},\rho_{_A}]$ has the
form~\cite{McLerV1,McLerV2,McLerV3,JeonV1},
\begin{eqnarray}
W_{_{A}}[x_{_A},\rho_{_A}] = \exp\left( -\int d^2 \x \left[
\frac{\rho_{_A}^a(\x)\rho_{_A}^a(\x)}{2\mu_{_A}^2}
-
d_{abc}\,\frac{\rho_{_A}^a(\x) \rho_{_A}^b(\x) \rho_{_A}^c(\x)}{\kappa_{_A}}\right]
\right) \; ,
\label{eq:GW}
\end{eqnarray}
where~\footnote{Note that the number density $\rho_A$ is related to the charge density ${\tilde \rho}_A$ by ${\tilde \rho}_A =g \rho_A$. 
For a discussion of different conventions, see Ref.~\cite{KrasnNV1}.} (see eq.~(\ref{eq:Gauss-corr}))
\begin{equation}
\mu_{_A}^2 = \frac{A}{ 2\pi R^2} \,\,\,\,\,\,{\rm and}\,\,\,\,\,\,
\kappa_{_A} = \frac{A^2 N}{\pi^2 R^4 } \; .
\label{eq:mua}
\end{equation}
The cubic term in the weight functional is parametrically suppressed
by $A^{1/6}$, and since we are working in the limit that $A^{1/6} \gg
1$, we will restrict ourselves in the rest of the discussion to the
Gaussian term alone.

The lower limit of the stated kinematical domain in rapidity follows
from the requirement that $x_{_A} \ll A^{-1/3}$, which ensures that
the small $x$ partons in the nucleus couple coherently to all the
color charges present along the $z$ direction. The upper bound follows
from the constraint that quantum corrections with a leading logarithm
in $x$ are small, namely, $\alpha_s \ln(1/x_{_A}) \ll 1$. This
condition combined with the large nucleus condition, $\alpha_s^2
A^{1/3} \gg 1$, leads to the stated upper bound. In this kinematical
domain in rapidity, for large nuclei, the model of Gaussian
correlations of classical ``valence" charges in the nuclear
wave-function -- the McLerran-Venugopalan model --  is valid.

The MV model is thus a plausible model at moderately small values of
$x$ ($x\sim 10^{-2}$), where small $x$ quantum evolution effects are
not yet large. It gives reasonable results for the initial conditions
in heavy ion collisions at RHIC
~\cite{KrasnNV1,KrasnNV2,KharzN1,KharzLN2,HiranN1} and the Cronin
effect in Deuteron-Gold
collisions~\cite{DumitJ1,DumitJ2,GelisJ3,KharzLM1,BaierKW1,JalilNV1,AlbacAKSW1},
also at RHIC. In this Gaussian approximation, one can compute the
2-point, 3-point and 4-point correlators in closed
form~\cite{JalilKMW1,McLerV4,GelisP1,KovneW1,Fujii1,FujiiM1,Tuchi1,BlaizGV2}
-- see Appendix A of ref.~\cite{BlaizGV2} for the explicit expressions
for arbitrary $N$. We quote here the large $N$ limits for these
correlators because only these are used in the discussion of the
energy evolution of correlators\footnote{The numerical evaluation of
the $Q\overline{Q}$ cross-section at finite $N$ would be much more
complicated and time consuming, because of the very complicated
structure of the exact 4-point function (see \cite{BlaizGV2} for the
complete formula).}~:
\begin{eqnarray}
{\rm tr}\big<{\wt U}(\x_\perp)t^a {\wt U}^\dagger(\y_\perp)
{\wt U}(\u_\perp)t^a {\wt U}^\dagger(\v_\perp)\big>  &\equiv& C(\x_\perp,\y_\perp;\u_\perp,\v_\perp)\nonumber \\
&\empile{=}\above{N\to \infty}&
\frac{N^2}{2}
e^{-\frac{N}{4}\mu_{_{A}}^2\left[\Gamma(\x_\perp-\v_\perp)
+\Gamma(\y_\perp-\u_\perp)\right]}\; ,
\nonumber\\
&&
\label{eq:largeN}
\end{eqnarray}
where we denote 
\begin{equation}
\Gamma(\y_\perp-\u_\perp) \equiv g^4 \int d^2\z_\perp \Big[G_0(\y_\perp - \z_\perp)-G_0(\z_\perp-\u_\perp)\Big]^2\; ,
\label{eq:Gamma}
\end{equation}
with 
\begin{equation}
G_0(\x_\perp-\z_\perp)\equiv \int\frac{d^2\k_\perp}{(2\pi)^2}\;
\frac{e^{i\k_\perp\cdot(\x_\perp-\z_\perp)}}{\k_\perp^2} \; .
\label{eq:2Dpropagator}
\end{equation}
The other two correlators are obtained as special cases of the
argument of the 4-point function $C$
\begin{eqnarray}
{\rm tr}\left<{\wt U}(\x_\perp)t^a {\wt U}^\dagger(\y_\perp)
t^b U^{\dagger ab}(\u_\perp)
\right>&=&C(\x_\perp,\y_\perp;\u_\perp,\u_\perp)\nonumber\\
{\rm tr}\left<
t^b U^{ba}(\x_\perp) t^{b^\prime}
U^{\dagger a b^\prime}(\u_\perp)
\right>
&=&
C(\x_\perp,\x_\perp;\u_\perp,\u_\perp)\; .
\label{eq:2C3C}
\end{eqnarray}
Thus in the large $N$ limit, all the correlators that we need in order
to compute the cross-section in eq.~(\ref{eq:cross-section}) can be
obtained from the quantity $\Gamma(\x_\perp-\y_\perp)$, defined by
eqs.~(\ref{eq:Gamma}) and ~(\ref{eq:2Dpropagator}).

When performing the Fourier transforms specified by
eqs.~(\ref{eq:corr-nucleus}) to obtain the $\phi_{_A}$'s that enter in
the cross-section, we note that in the large $N$ limit we have
\begin{equation}
\phi_{_A}^{q\bar{q},q\bar{q}}(\l_\perp;\k_\perp,\k_\perp^\prime)
=(2\pi)^2\delta(\k_\perp-\k_\perp^\prime)\;
\phi_{_A}^{q\bar{q},g}(\l_\perp;\k_\perp)\; .
\end{equation}
This relation makes the sum-rule that relates the 3- and 4-point
functions completely obvious. Moreover, it makes the numerical
integration simpler since it eliminates the integration over
$\k^\prime$. For the purpose of numerical integrations, in the large N limit, the 
$Q\overline{Q}$ cross-section can be efficiently written as 
\begin{eqnarray}
&&\frac{d\sigma}{d^2\p_\perp d^2\q_\perp dy_p dy_q}=
\frac{\alpha_s^2 N}{8\pi^4 (N^2-1)}
\int\limits_{\k_{1\perp},\k_{2\perp}}\!\!\!
\frac{\delta(\p_\perp+\q_\perp-\k_{1\perp}-\k_{2\perp})}
{\k_{1\perp}^2 \k_{2\perp}^2}\,\varphi_p(\k_{1\perp})
\nonumber\\
&&\qquad\qquad\!\!\!\!\!
\times
\int_{\k_\perp}\phi_{_A}^{q\bar{q},g}(\k_{2\perp};\k_\perp)
\;\Bigg\{
{\rm tr}_{\rm d}
\Big[(\slq\!+\!m)T_{q\bar{q}}(\slp\!-\!m)
\gamma^0 T_{q\bar{q}}^{\dagger}\gamma^0\Big]
\nonumber\\
&&\qquad\qquad\qquad\qquad\qquad\qquad
+
{\rm tr}_{\rm d}
\Big[(\slq\!+\!m)T_{q\bar{q}}(\slp\!-\!m)
\gamma^0 T_{g}^{\dagger}\gamma^0 + {\it h.c.}\Big]
\nonumber\\
&&\qquad\qquad\qquad\qquad\qquad\qquad
+{\rm tr}_{\rm d}
\Big[(\slq\!+\!m)T_{g}(\slp\!-\!m)\gamma^0 T_{g}^{\dagger}\gamma^0\Big]
\Bigg\}
\; ,
\label{eq:cross-section-1}
\end{eqnarray}
where we have used the sum rule relating the 2- and
3-point functions to express the result in terms of the 3-point function alone.
This expression is numerically very useful because exact cancellations that occur
at small $\k_{1\perp}$ among the three terms of
eq.~(\ref{eq:cross-section}) are a consequence of the sum rules satisfied by the three $\phi_{_A}$'s.  
This is difficult to ensure numerically and rewriting the cross-section by building in the sum rules, as in
eq.~(\ref{eq:cross-section-1}), is an efficient way to achieve it. 

\subsection{Energy evolution of correlators via the BK equation}
Recall that the previous results are limited to the regime where
$\alpha_s \ln(1/x_{_A}) \ll 1$. They therefore properly include
multiple scattering effects which exist due to the large density of
scatterers.  However, they do not include the quantum effects arising
from small-$x$ evolution, that generate leading twist
shadowing.

The assumption of Gaussian correlations breaks down when
$\alpha_s \ln(1/x)\sim 1$~\cite{AyalaJMV1,AyalaJMV2}. The JIMWLK
renormalization group
equations~\cite{JalilKLW1,JalilKLW2,JalilKLW3,JalilKLW4,KovneM1,
KovneMW3,Balit1,JalilKMW1,IancuLM1,IancuLM2,FerreILM1} incorporate, in
the CGC framework, these large quantum corrections in the $x$
evolution of $W[x_{_A},\rho_{_A}]$. Consider for instance the
correlator of two fundamental Wilson lines, $\langle {\wt
U}(\x_\perp)\,{\wt U}^\dagger(\y_\perp)\rangle$, where the brackets
denote the average with the nuclear weight functional
$W[x_{_A},\rho_{_A}]$. This correlator is directly proportional to the
total $q{\bar q}A$ cross-section in deeply inelastic
scattering~\cite{McLerV4,GelisP1} and appears as well in the single
quark production cross-sections~\footnote{The single quark
cross-section can be obtained by integrating over the kinematic
variables of one of the quarks in the
pair~\cite{BlaizGV2,FujiiGV1}. Using the sum rule in
eq.~(\ref{eq:int-2}), one can show that the correlators that
contribute to single quark production are the 2-point correlators
$\phi_{_A}^{g,g}(\l_\perp)$ and $\phi_{_A}^{q,q}(\l_\perp)$, and the
3-point correlator $\phi_{_A}^{q\bar{q},g}(\l_\perp;\k_\perp)$. The
correlator $\phi_{_{A}}^{q,q}$ is the Fourier transform of $\langle
{\wt U}(\x_\perp)\,{\wt U}^\dagger(\y_\perp)\rangle$.}.  It satisfies
the renormalization group equation,
\begin{eqnarray}
&&-\frac{d}{d\ln\left({x_0/x}\right)}
{\rm tr}\left<
{\wt U}^\dagger({\x_\perp}) 
{\wt U}({\y_\perp})
\right>
 = 
\frac{\alpha_s}{2\pi^2}\,
\int d^2 \z_\perp\,
\frac{(\x_\perp -\y_\perp)^2}
{(\x_\perp-\z_\perp)^2\,(\z_\perp-\y_\perp)^2}\nonumber \\
&&\qquad\times \left< N\, 
{\rm tr}[ 
{\wt U}^\dagger({\x_\perp}) 
{\wt U}({\y_\perp})] 
- 
{\rm tr}\,[
{\wt U}^\dagger({\x_\perp}) 
{\wt U}({\z_\perp})
]
\,{\rm tr}\,[
{\wt U}^\dagger({\z_\perp}) 
{\wt U}({\y_\perp})]\right>\; .
\nonumber\\
&&
\label{eq:Balitsky}
\end{eqnarray}
This equation was first derived by Balitsky~\cite{Balit1} and
subsequently discussed in the CGC framework in
Refs.~\cite{JalilKLW1,JalilKLW2,JalilKLW3,JalilKLW4,KovneM1,
KovneMW3,JalilKMW1,IancuLM1,IancuLM2,FerreILM1}.  It cannot be solved
in closed form since the 2-point function depends on the 4-point
function,, and so on. However, in the large $N$ {\sl and} large $A$
($\alpha_s^2 A^{1/3} \gg 1$) limit, the correlator of the product of
traces of two pairs of Wilson lines factorizes into the product of the
correlators of traces of pairs of Wilson lines:
\begin{eqnarray}
&& \left< \,{\rm tr}\,[{\wt U}^\dagger({\x_\perp}) {\wt U}({\z_\perp})]
\,\,{\rm tr}\,[{\wt U}^\dagger({\z_\perp}) {\wt U}({\y_\perp})]\,\right>
\nonumber \\
&&\qquad=
\left< \,{\rm tr}\,[{\wt U}^\dagger({\x_\perp}) {\wt U}({\z_\perp})]\,\right>
\,
\left<\,{\rm tr}\,[{\wt U}^\dagger({\z_\perp}) {\wt U}({\y_\perp})]\,\right> 
\; ,
\label{eq:Factorization}
\end{eqnarray}
to leading order in a $1/N$ expansion. With this factorization, eq.~(\ref{eq:Balitsky}) becomes a closed form 
equation, called the Balitsky-Kovchegov (BK)
equation~\cite{Balit1,Kovch3}.  While this equation has not been
solved analytically (see however footnote \ref{foot:noteBK}), as
discussed in the introduction, this equation has been solved
numerically by several groups for both the fixed and running coupling
cases~\cite{AlbacAKSW1,Braun4,Lubli1,LevinL1,GolecMS1,GotsmKLMN1,RummuW1,AlbacAMSW1}. For
inclusive gluon production, where $k_\perp$-factorization
holds~\cite{KharzKT1}, the energy evolution of the two point
correlator $\phi_{_{A}}^{g,g}$ has been studied using the BK evolution
equation~\cite{AlbacAKSW1}. These studies show the same qualitative
behavior as those seen in the remarkable measurements~\cite{Arsena1}
in Deuteron-Gold collisions at RHIC of the disappearance of the Cronin
peak with rapidity, and of the reversal of the centrality dependence
of this effect from central to forward rapidities.

These studies, with the BK equation, of the energy dependence of
inclusive gluon production can be extended to quark pair
production. This is because the factorization in
eq.~(\ref{eq:Factorization}), essential to the derivation of the BK
equation, requires that correlations between color charges be
Gaussian.  This may seem odd at first glance because we just argued
that the MV model fails when $\alpha_s\ln(1/x)\approx 1$. The
resolution of this apparent paradox is that the correlations are not
necessarily local anymore, namely,
\begin{equation}
\langle\rho_{_{A},a}(\x_\perp)\rho_{_A,a^\prime}(\y_\perp)\rangle
=
\delta_{aa^\prime}{\bar \mu}_{_{A}}^2(x,\x_\perp-\y_\perp)\; .
\label{eq:nonlocalGauss-corr}
\end{equation}
The 2-, 3- and 4-point correlators are computed as in
eqs.~(\ref{eq:largeN}-\ref{eq:2C3C}), with eq.~(\ref{eq:Gamma})
replaced by
\begin{eqnarray}
\mu_{_A}^2\,\Gamma(x,\y_\perp-\u_\perp)
&\equiv&
g^4\,\int d^2 \z_\perp d^2 \r_\perp\;
 {\bar \mu}_{_{A}}^2(x,\z_\perp-\r_\perp) \nonumber \\
&&\qquad\times 
\left( G_0(\y_\perp - \z_\perp) - G_0(\u_\perp - \z_\perp)\right)
\nonumber\\
&&\qquad\times 
\left( G_0(\r_\perp-\y_\perp) - G_0(\r_\perp -\u_\perp)\right) \, .
\label{eq:BKapprox}
\end{eqnarray}
The $x$-dependent l.h.s. of this equation can be determined directly
by solving the BK equation for $\langle {\wt U}(\y_\perp)\,{\wt
U}^\dagger(\u_\perp)\rangle$. Indeed, we have
\begin{equation}
\frac{N}{4}\mu_{_A}^2\,\Gamma(x,\y_\perp-\u_\perp)
=
-\ln \langle {\wt U}(\y_\perp)\,{\wt
U}^\dagger(\u_\perp)\rangle\; .
\end{equation}
The r.h.s. of this equation is obtained by solving the BK equation in
eqs.~(\ref{eq:Balitsky}) and (\ref{eq:Factorization}) with initial
conditions given by the MV model. The l.h.s., thus determined, specifies
the values of the correlators in eq.~(\ref{eq:largeN}) and
eq.~(\ref{eq:2C3C}).  Therefore, in the large $A$ and large $N$ limit,
we can compute the energy dependence of all the correlators involved
in pair production. This will enable us to study the effects of both
multiple scattering and quantum evolution (shadowing) on the pair
production cross-section. In the next section, we will discuss our
results obtained in this approach.

\section{Results on pair production}
\label{sec:results}
In this section, we will discuss results of our computation of the
pair production cross-section in eq.~(\ref{eq:pair-cross-section})
using the results for the correlators in the MV and BK models. The
former includes multiple scattering effects but does not include small
$x$ evolution. The latter includes both; it leads to a
non-trivial $A$ dependence, often called leading twist shadowing, that
goes away only logarithmically with momentum. We would like to
understand the dependence of pair production on the pair mass, the
pair momentum, the quark mass and the rapidity of the pair. We would like to know how
it depends on the saturation scale $Q_s$. The saturation scale is a
measure of the parton density in the system and depends on both $x$
and $A$. In the first subsection, we will present results in the MV
model to study the impact of the multiple scatterings alone on these
quantities. In the second subsection, we will consider how small $x$
quantum evolution {\it a la BK} modifies these results. As discussed
previously, the MV results may be more relevant at central rapidities
at RHIC; quantum evolution effects may be more relevant in forward
Deuteron-Gold studies at RHIC and already for central rapidities at the LHC.

\subsection{MV model:  multiple scattering effects}
\subsubsection{$M$ and $\P_\perp$ dependence}
\begin{figure}[htbp]
\begin{center}
\resizebox*{8cm}{!}{\includegraphics{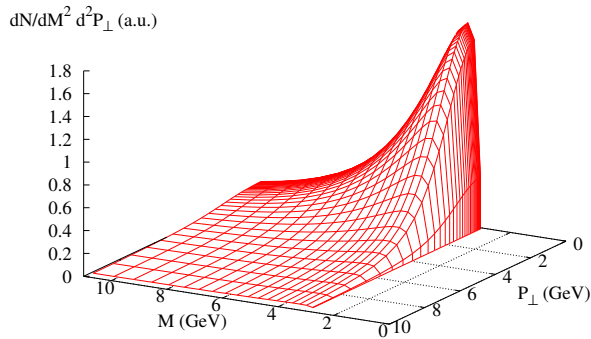}}
\end{center}
\caption{\label{fig:pairspectrum} Pair cross-section (in arbitrary units) in the MV model,
as a function of $\P_\perp$ and $M$.}
\end{figure}
In fig.~\ref{fig:pairspectrum}, for illustration, we plot the
cross-section for the quark pair production as a function of
$\P_\perp$ and $M$.  We have chosen here a quark mass of $m$=1.5 GeV
and the saturation scale~\footnote{\label{foot:muQs}The dipole-hadron
cross-section measured in DIS can be parameterized in terms of the
saturation scale {\it a la}
Golec-Biernat--Wusthoff~\cite{GolecW1,GolecW2}. It is related to the
scale $\mu_A^2$ in the MV model by the relation $g^4\mu_A^2 = 4\pi
Q_s^2/C_F \ln(Q_s^2/\Lambda_{\rm QCD}^2)$, where $C_F$ is the Casimir
in the fundamental representation.} to be $Q_s^2$=2
GeV$^2$. Unsurprisingly, it is peaked at small values of $\P_\perp$
and just above the threshold in the invariant mass $M$ of the pairs.

Examining the behavior of the
differential pair cross-section at large $M$ and fixed $P_\perp$, or
conversely at large $P_\perp$ and fixed $M$, we obtain the 
asymptotic forms, 
\begin{eqnarray}
&&
\frac{dN}{dM^2 d^2\P_\perp dY}
\empile{\sim}\above{{M\to\infty}\atop{P_\perp={\rm const}}}
\frac{\ln^2(M^2)}{M^4}\; ,
\nonumber\\
&&
\frac{dN}{dM^2 d^2\P_\perp dY}
\empile{\sim}\above{{P_\perp\to\infty}\atop{M={\rm const}}}
\frac{\ln(P_\perp^2)}{P_\perp^4}\; .
\end{eqnarray}
 The $1/P_\perp^4$ and the $1/M^4$ behavior, in the stated limits is as one would expect in
perturbative QCD (pQCD). Further, the logarithmic prefactors can be related to the collinear
logarithms of pQCD which, at leading log order and beyond, are absorbed in the
hadronic structure functions. 

The occurrence of these logarithms in the CGC framework and their relation to the collinear logs of pQCD can be
understood as follows. The CGC formulas obtained in \cite{GelisV1}
and \cite{BlaizGV2} correspond to color sources from the proton
and the nucleus  that radiate gluons which fuse into a quark-antiquark pair.
\begin{figure}[htbp]
\begin{center}
\resizebox*{7cm}{!}{\includegraphics{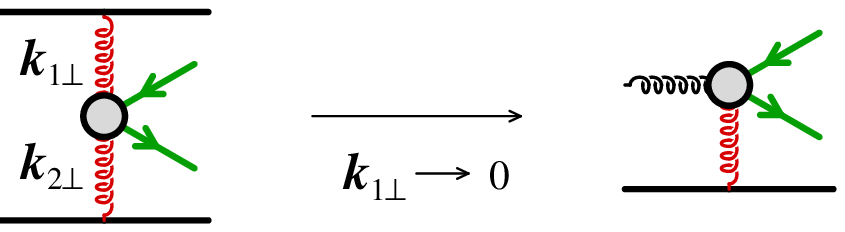}}
\vglue 4mm
\resizebox*{7cm}{!}{\includegraphics{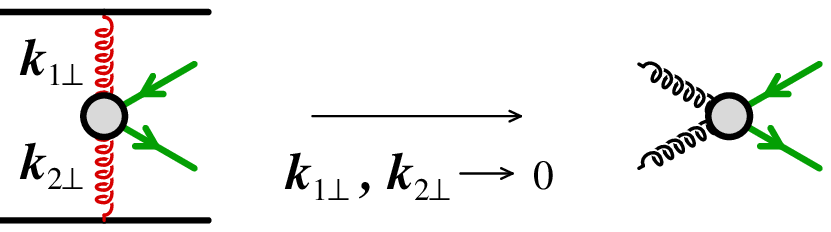}}
\end{center}
\caption{\label{fig:coll-limit} Diagrams in the framework of collinear
factorization obtained when one (top), or both (bottom), of
the transverse momentum transfer to the produced pair from the 
proton and nucleus becomes small.}
\end{figure}
In the limit where both the gluons from the proton and from the
nucleus have a small transverse momentum, the CGC formulas reduce to
the process $gg\to Q\overline{Q}$ (at leading order) in the collinear
factorization framework. Moreover, the integration over the transverse
momenta of these gluons produce logarithms that can be interpreted as
the first power of the collinear logarithms  resummed by the
DGLAP equation. Because the large $M$ limit is accessible at leading
order in collinear factorization, it is natural to expect two powers
of $\ln(M^2)$ in our formulas corresponding to the limit
$|\k_{1\perp}|$, $|\k_{2\perp}|\rightarrow 0$. This is shown in the
lower diagram of fig.~\ref{fig:coll-limit}. In contrast, the limit of
large $P_\perp$ of the pair can only be studied at next-to-leading
order in the collinear factorization framework. In our formula, as
shown in the upper diagram of fig.~\ref{fig:coll-limit}, this
corresponds to the limit where only one of the two gluons has a small
transverse momentum ($|\k_{1\perp}|\rightarrow 0$). Hence the single
power of $\ln(P_\perp)$. The relation of higher orders in the
collinear factorization framework for pair production to the $k_\perp$
factorization framework was explored previously in
refs.~\cite{CatanCH1,CatanCH2}.

\subsubsection{Breaking of $k_\perp$-factorization}
\begin{figure}[htbp]
\begin{center}
\hfill
\resizebox*{5.5cm}{!}{\includegraphics{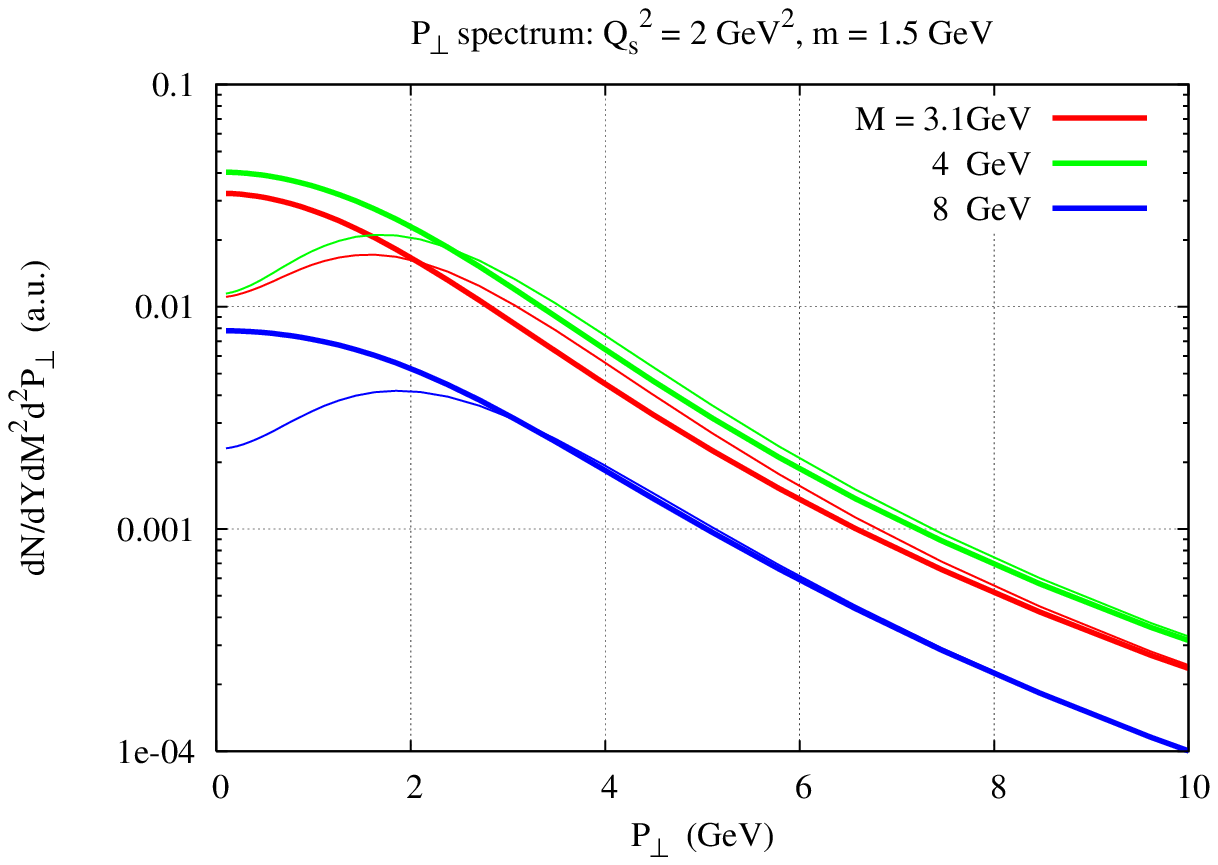}}
\hfill
\resizebox*{5.5cm}{!}{\includegraphics{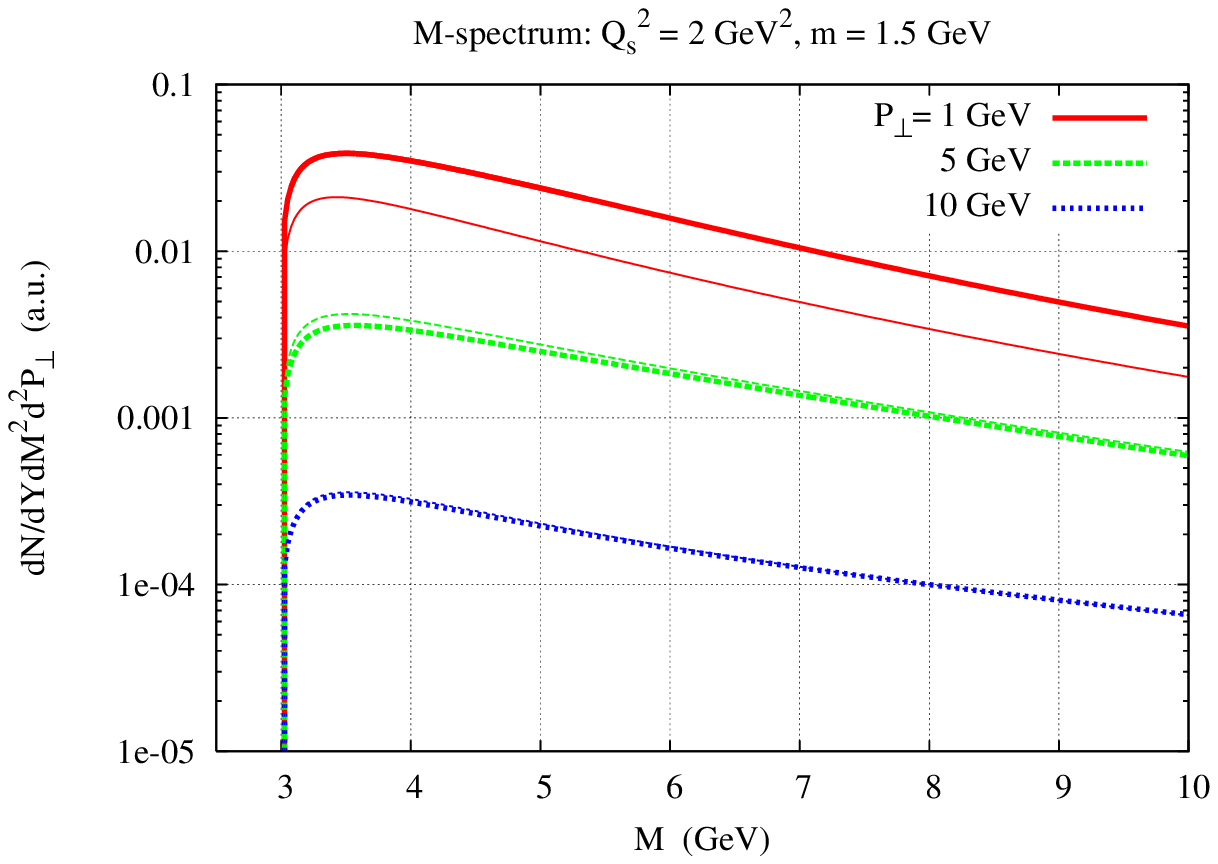}}
\hfill
\end{center}
\caption{\label{fig:exactvsktfact} Pair cross-section as a function of
$\P_\perp$ for fixed $M$ (left).  Pair cross-section as a function of
$M$ for fixed $\P_\perp$ (right).  The results in the
$\k_\perp$-factorized approximation are shown with thin lines.}
\end{figure}

Another important issue, previously studied for the production of single
quarks in \cite{FujiiGV1}, is that of $k_\perp$-factorization.  This
is illustrated in figure~\ref{fig:exactvsktfact}, where we display the
cross-section versus $P_\perp$ for fixed $M$ (left) and versus $M$ for
fixed $P_\perp$ (right). In these figures, we compare the exact values
from eq.~(\ref{eq:pair-cross-section}) (using
eq.~(\ref{eq:cross-section}) and eq.~(\ref{eq:corr-nucleus}) to the
same expression in the $k_\perp$-factorized approximation for pair
production. 
The $k_\perp$-factorized result is obtained by replacing
$\phi_{_A}^{g,q\bar{q}}$ and $\phi_{_A}^{q\bar{q},q\bar{q}}$ in
eq.~(\ref{eq:cross-section}) by
 \begin{eqnarray}
&&
\phi_{_A}^{g,q\bar{q}}(\k_{2\perp};\k_\perp)_{\rm fact}
\equiv
\phi_{_A}^{g,g}(k_{2\perp})
(2\pi)^2\left[\delta(\k_\perp)\!+\!\delta(\k_\perp\!-\!\k_{2\perp})\right]
\nonumber\\
&&
\phi_{_A}^{q\bar{q},q\bar{q}}(\k_{2\perp};\k_\perp,\k_\perp^\prime)_{\rm fact}
\equiv
\phi_{_A}^{g,g}(k_{2\perp})
\nonumber\\
&&\qquad\qquad\times
(2\pi)^4\Big[
\frac{C_{_{F}}}{N} \left(
\delta(\k_\perp-\k_{2\perp})\delta(\k_\perp^\prime-\k_{2\perp})
+
\delta(\k_\perp)\delta(\k_\perp^\prime)
\right)
\nonumber\\
&&\qquad\qquad\qquad
+\frac{1}{2N^2}
\left(
\delta(\k_\perp-\k_{2\perp})\delta(\k_\perp^\prime)
+
\delta(\k_\perp)\delta(\k_\perp^\prime-\k_{2\perp})
\right)
\Big]\; .\nonumber\\
&&
\end{eqnarray}
In other words, the limit of $k_\perp$-factorization is obtained from
``nuclear distributions'' very similar to the leading twist formulas
in eqs.~(\ref{eq:LT}), except that the leading twist 2-point function
is replaced by the ``all twist" 2-point function. This means that some of the rescattering
corrections, but not all of them, can be included in the approximation
of $k_\perp$-factorization.

\begin{figure}[htbp]
\begin{center}
\resizebox*{8cm}{!}{\includegraphics{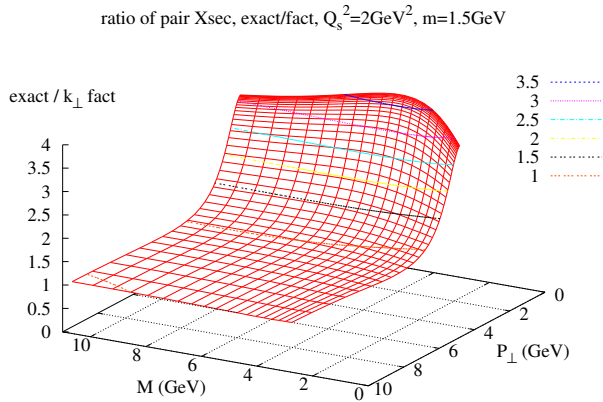}}
\end{center}
\caption{\label{fig:kt-ratio}Ratio of the pair cross-section of the
full result to the $k_\perp$-factorization result as a function of
$\P_\perp$ and $M$.}
\end{figure}

The dependence of the ratio of the exact result to the
$k_\perp$-factorized result, as a function of $\P_\perp$ and $M$, is 
nicely seen in the 3d plot of fig.~\ref{fig:kt-ratio}.
At any fixed $M$, for large $P_\perp$, the exact
cross-section and the $k_\perp$-factorized approximation become
identical. (See also the left plot of figure \ref{fig:exactvsktfact}.) This
is because, in this limit, the quark and the anti-quark 
become collinear with each of them having a very large transverse
momentum. The quark-antiquark pair then scatters off the medium as a gluon would; as in the latter case, this leads to 
$k_\perp$ factorization.

On the contrary, we observe in the right
plot of figure \ref{fig:exactvsktfact} (fixed $P_\perp$ and large $M$), the
exact cross-section and the $k_\perp$-factorized approximation are not 
identical if the fixed value of the transverse momentum of the
pair, $P_\perp$, is of the order of $Q_s$ or smaller. This is because 
any pair configuration with a small total $P_\perp$ is very
sensitive to rescatterings;  even a small number of additional
rescatterings, regardless of the pair invariant mass, may significantly change the transverse momentum of the
pair.

Turning now to fixed invariant mass and smaller transverse momentum
$P_\perp\lesssim Q_s$, we note (see left panel of
fig.~\ref{fig:exactvsktfact}) a qualitative change in the behavior of the
cross-section due to multiple scattering, high parton density
effects. In the $k_\perp$-factorized approximation, the pair
cross-section shows a bump at $\P_\perp \sim Q_s$. Further it is
suppressed relative to the exact result for $\P_\perp \leq Q_s$.  This
suppression occurs because $k_\perp$-factorization requires that only
the quark or the anti-quark -- and {\it not both} -- scatters off the
nucleus.  The typical transverse momentum taken from the (dilute)
proton is rather small; the transverse momentum of the pair is
therefore approximately equal to the transverse momentum exchanged
during these scatterings on the nucleus--of order $Q_s$.  Thus, in the
$k_\perp$-factorized approximation, the pair is less likely to have a
total momentum $\P_\perp$ smaller than $Q_s$.  In contrast, in the
exact expression within the MV model, {\it both} the quark and the
anti-quark get multiply scattered in the target nucleus. Therefore
their net momentum can still be smaller than $Q_s$. This explains the
absence of a bump structure in the exact expression for the $\P_\perp$
distribution of the quark--anti-quark pair. The size of the
factorization breaking in this low $P_\perp$ region can be as large as
a factor 3.

\begin{figure}[htbp]
 \begin{center}
 \hfill
 \resizebox*{5.5cm}{!}{\includegraphics{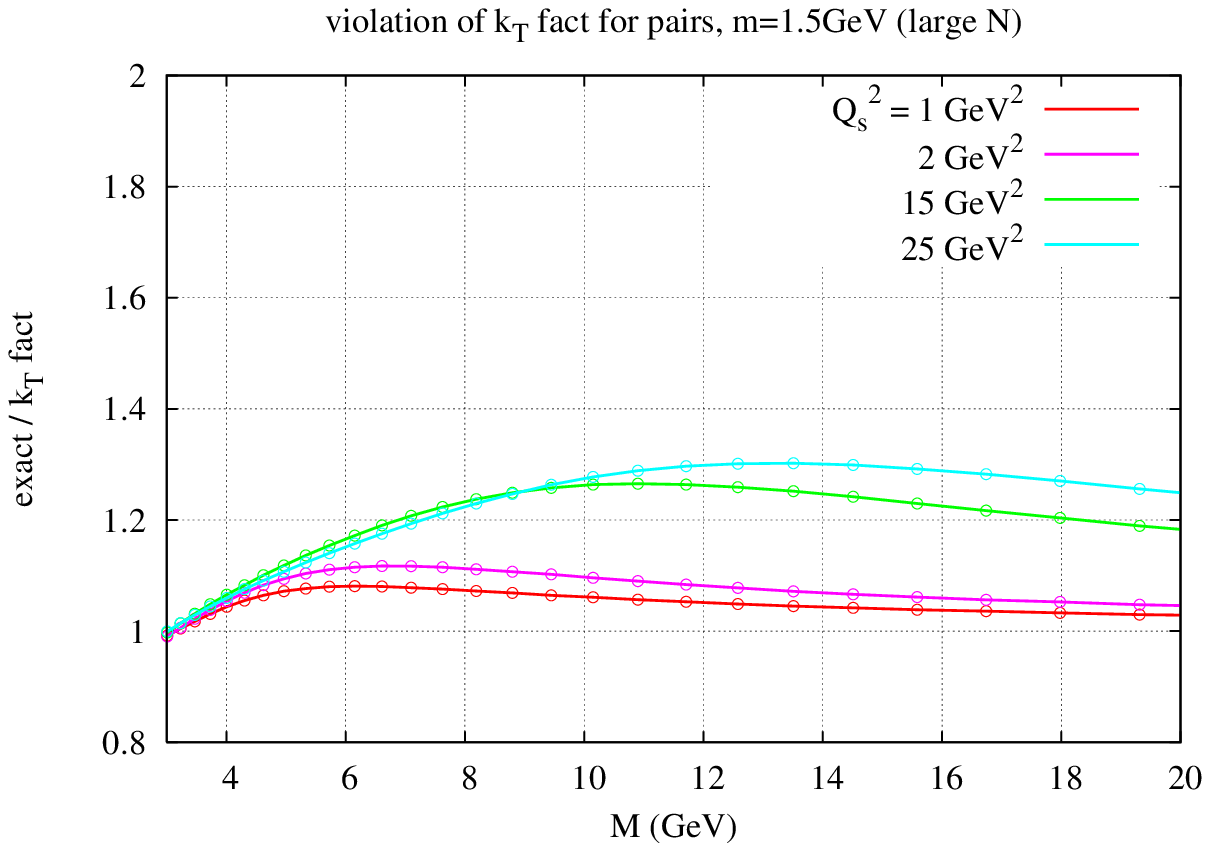}}
 \hfill
 \resizebox*{5.5cm}{!}{\includegraphics{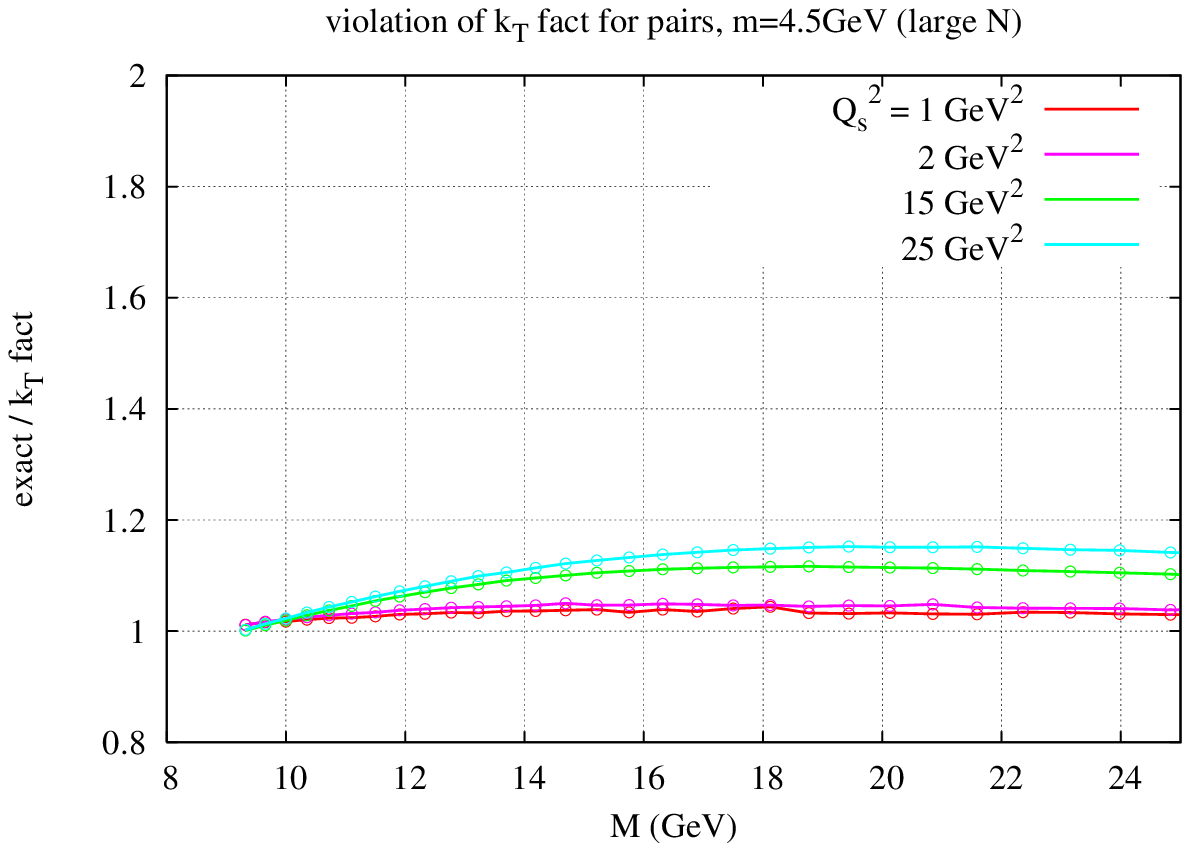}}
 \hfill
 \end{center}
 \caption{\label{fig:pairspectrum2} Ratio of the pair yields
 $dN/dYdM^2$ from the exact formula and from the $k_\perp$-factorized
 approximation as a function of $M$ for $m$=1.5~GeV (left) and
 4.5~GeV (right).}
 \end{figure}
When we integrate the cross-section over the total momentum
$\P_\perp$, the magnitude of the factorization breaking, shown in
fig.~\ref{fig:pairspectrum2}, is about $10\%$ for $m$=1.5 GeV and
$Q_s^2$= 2 GeV$^2$. The latter is a typical value associated with
central RHIC collisions.  For the larger values of $Q_s$ that may be
accessed in forward Deuteron-Gold collisions at RHIC and the LHC, the
magnitude of the breaking is maximally $30\%$. Note that even for
large invariant masses, the ratio returns to unity very slowly.  This
again is due to the fact that in the low $\P_\perp$ region, which
contributes significantly to the integrated cross-section, the
relative magnitude of violation of $k_\perp$-factorization is constant
for large $M$. As for the case of single quark
production~\cite{FujiiGV1}, fig.~\ref{fig:pairspectrum2} shows that
the breaking of $k_\perp$-factorization grows with increasing $Q_s$
while it is systematically weaker for larger quark masses.

 Finally, we observe in fig.~\ref{fig:pairspectrum2} that, precisely
at the pair threshold, the ratio of the exact to the
$k_\perp$-factorized expression is unity.  As is clear from the left
panel of fig.~\ref{fig:kt-ratio}, this ratio is not unity for any
specific $P_\perp$ but is a property of the distribution integrated
over $P_\perp$. This observation seems related to the following fact:
at threshold, the quark and the antiquark are at rest in the rest
frame of the pair. When boosted to the lab frame, they have the same
momenta at threshold and are therefore collinear. A pair made of a
collinear quark and antiquark is very similar to an octet gluon. The
correlator of two Wilson lines satisfies a sum rule which ensures that
the momentum integrated distributions are unchanged even if the
momenta are redistributed by re-scattering. It is therefore plausible
that the integrated pair distribution at threshold, in analogy to the
gluon distribution, is insensitive to re-scatterings.

\subsubsection{Nuclear size ($Q_s^2$) dependence}
In the MV model, the nuclear size dependence arises only through the
radius\footnote{This dependence on the nuclear size via the area of
overlap between the two projectiles is trivial because translation
invariance of all the multiparton correlators in the transverse plane
is assumed.}  $R$ and the saturation scale $Q_s^2$. In the MV model,
the saturation scale $Q_s^2$ is independent of the
energy\footnote{Albeit often, in saturation models, the
Golec-Biernat--Wusthoff ansatz ($Q_s^2 = Q_0^2
\left(\frac{x_0}{x}\right)^\lambda$, with $\lambda \approx 0.3$, and
$Q_0^2 = 1$ GeV$^2$) is combined with the MV model. As discussed in
section 3, this is consistent only if the Gaussian weight functional
is non-local, as in the BK equation.} and also depends on the atomic
mass number roughly as $A^{1/3}$. Alternately, the scale $\mu_{_A}^2$
is used, which determines the number of color charges in the nucleus,
per unit of transverse area. It is proportional to $A^{1/3}$. The
relation between the two scales is specified in footnote
~\ref{foot:muQs}.
 
 \begin{figure}[htbp]
 \begin{center}
\hfill
 \resizebox*{5.5cm}{!}{\includegraphics{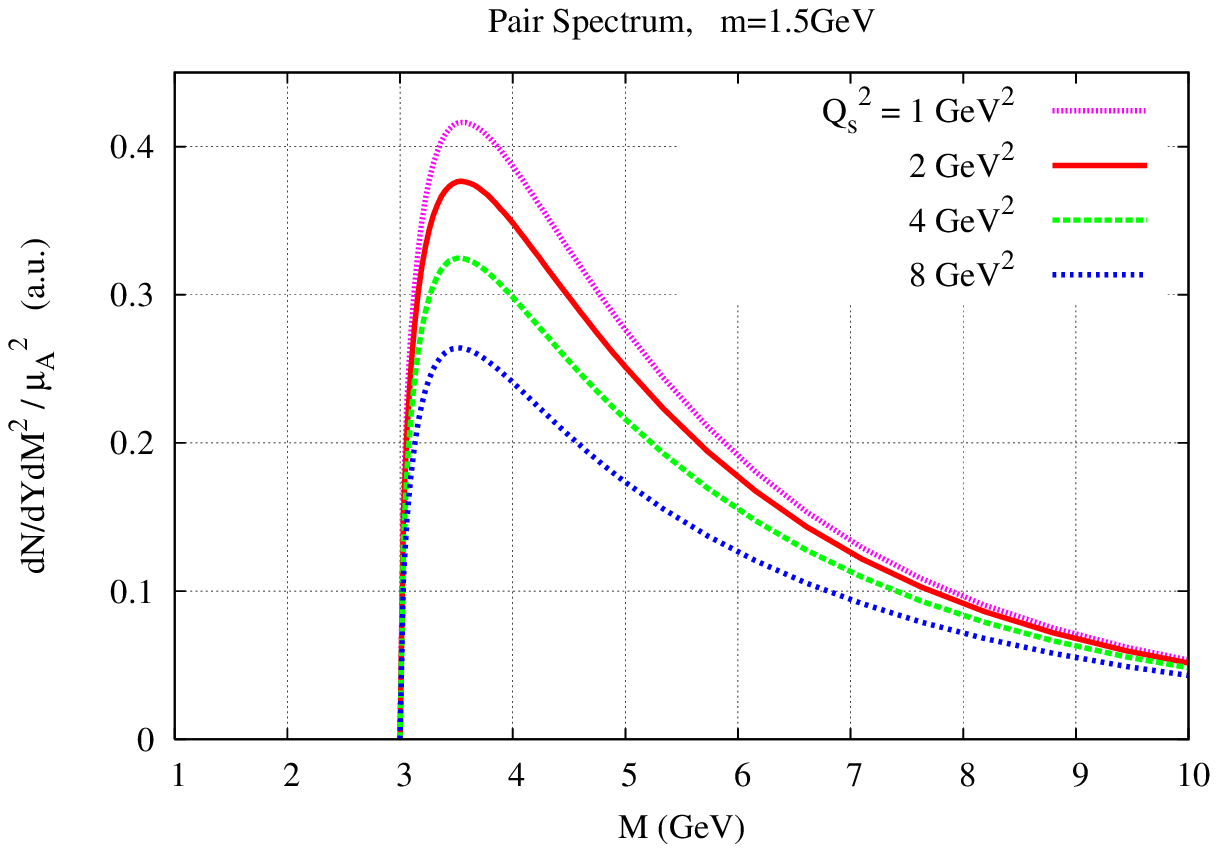}}
\hfill
\resizebox*{5.5cm}{!}{\includegraphics{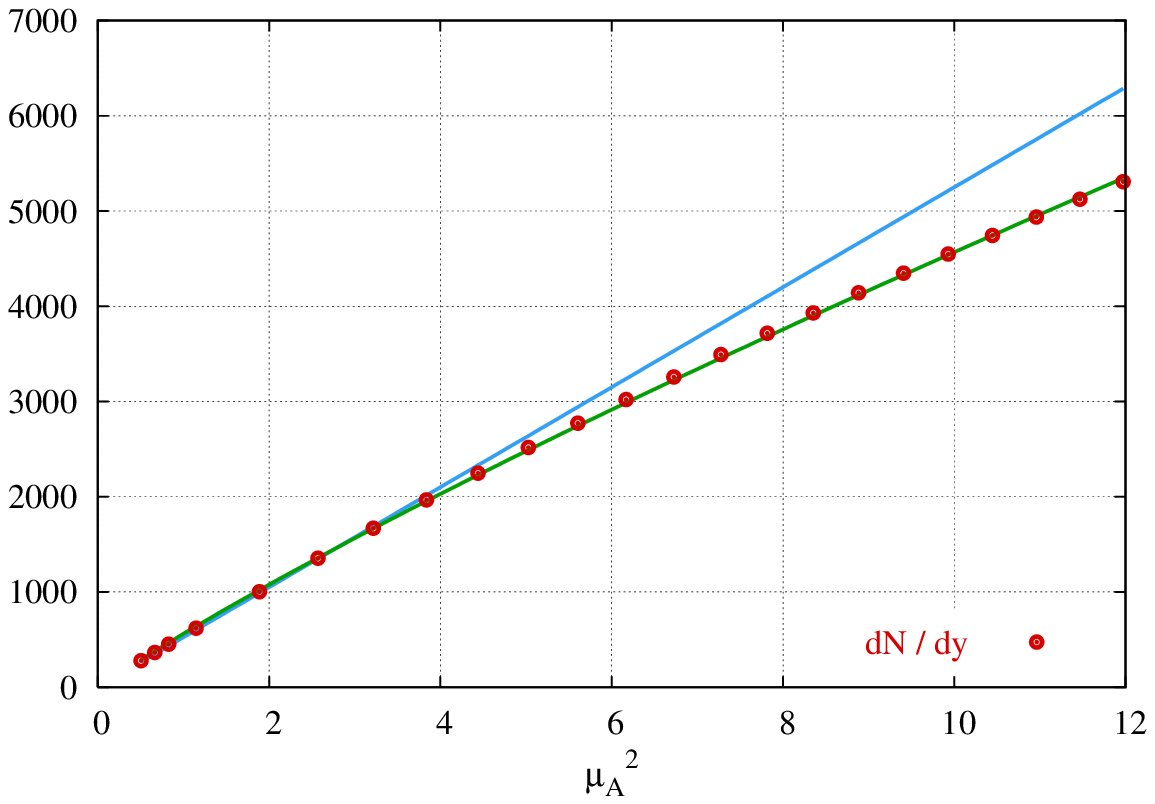}}
 \end{center}
 \caption{\label{fig:dndy-qs}Left: invariant mass spectrum of
 $Q\overline{Q}$ pairs for different values of the saturation
 scale. Right: integrated quark yield as a function of $\mu_{_A}^2\sim
 A^{1/3}$, the scale specifying the number of color charges per unit transverse area. Points: numerical evaluation of the cross-section. Solid
 line: fit by a function in $\mu_{_A}^2/\ln(\mu_{_A}^2)$.}
 \end{figure}
We first examine the integrated quark yield per unit rapidity (at
fixed impact parameter) as a function of the nuclear size. This
quantity is displayed in figure \ref{fig:dndy-qs}, in differential
form as a function of $M$ on the left plot and integrated as a
function of $\mu_{_A}^2\propto A^{1/3}$ on the right plot. Because one
naively expects, for heavy quarks, a scaling with the number of binary
collisions, namely $A^{1/3}$, the curves in the left plot have been
divided by a factor $\mu_{_A}^2$. The residual dependence on the
saturation scale is therefore a departure from the binary scaling
hypothesis. As one can see, when the saturation scale increases, the
differential yield decreases, albeit in a very moderate fashion (note
that the vertical axis for the left plot is a linear axis). When we
integrate these functions over the invariant mass in order to obtain
the total quark yield, we see that it is very close to a linear
function of $\mu_{_A}^2$, confirming the fact that the departure from
binary scaling is small. This departure is a small reduction of the
yield compared to binary scaling, and one can fit the dependence of
$dN/dY$ on $\mu_{_A}^2$ by a function that behaves like
$\mu_{_A}^2/\ln(\mu_{_A}^2)$.

Another quantity, whose dependence on the nuclear size is of 
phenomenological interest, is the nuclear modification factor
$R_{pA}$. It is  defined as the ratio between the yield in $pA$ collisions to
the yield in $pp$ collisions, normalized by the number of binary
collisions, 
\begin{equation}
R_{pA}\equiv \frac{1}{A^{1/3}}\;
\frac{
\left.\frac{dN}{dM^2 d^2\P_\perp dY}\right|_{pA}}
{\left.\frac{dN}{dM^2 d^2\P_\perp dY}\right|_{pp}}\; .
\label{eq:rpa-def}
\end{equation}
Indeed, this nuclear modification factor for $pA$ collisions is essential for establishing
 a baseline for ``normal nuclear suppression'' when 
 looking for potential quark gluon plasma suppression effects on the
yield of $J/\psi$'s in nucleus-nucleus collisions. 
\begin{figure}[htbp]
 \begin{center}
\hfill
 \resizebox*{5.5cm}{!}{\includegraphics{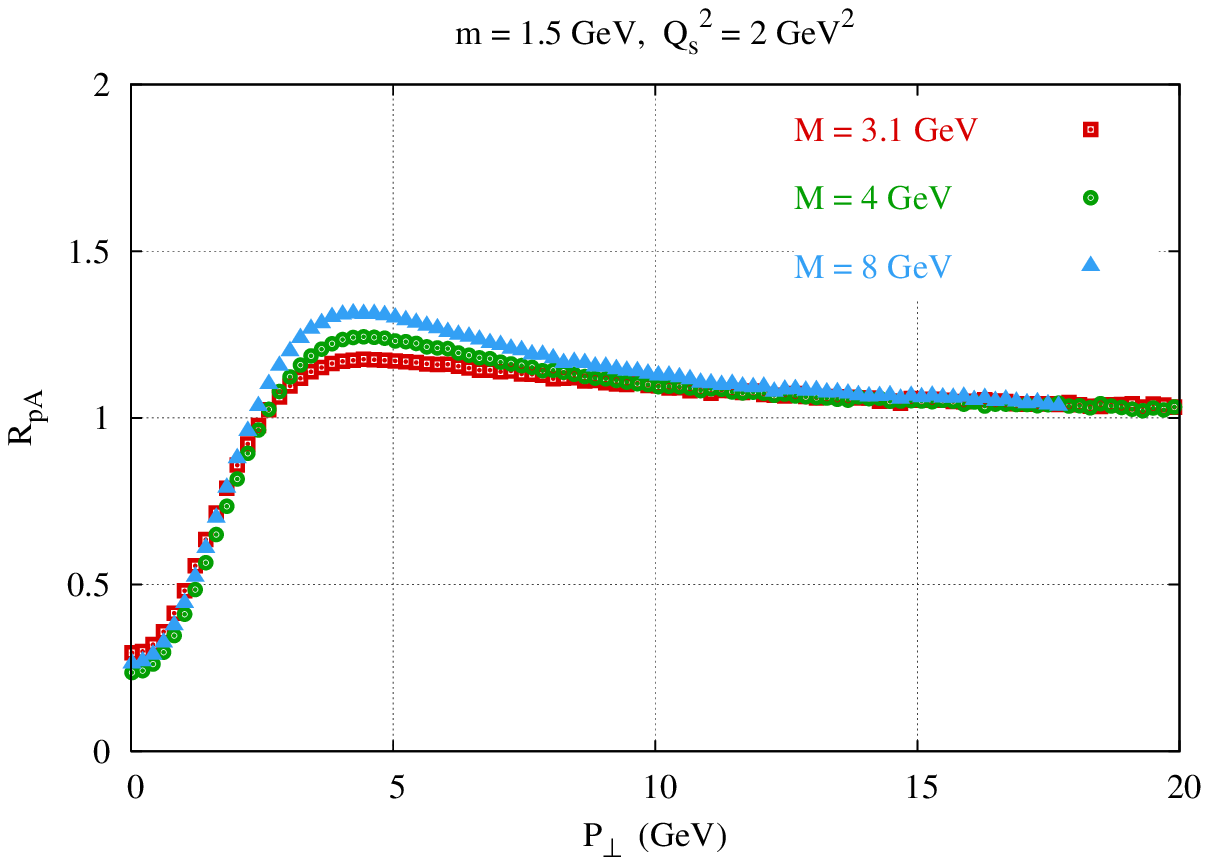}}
\hfill
\resizebox*{5.5cm}{!}{\includegraphics{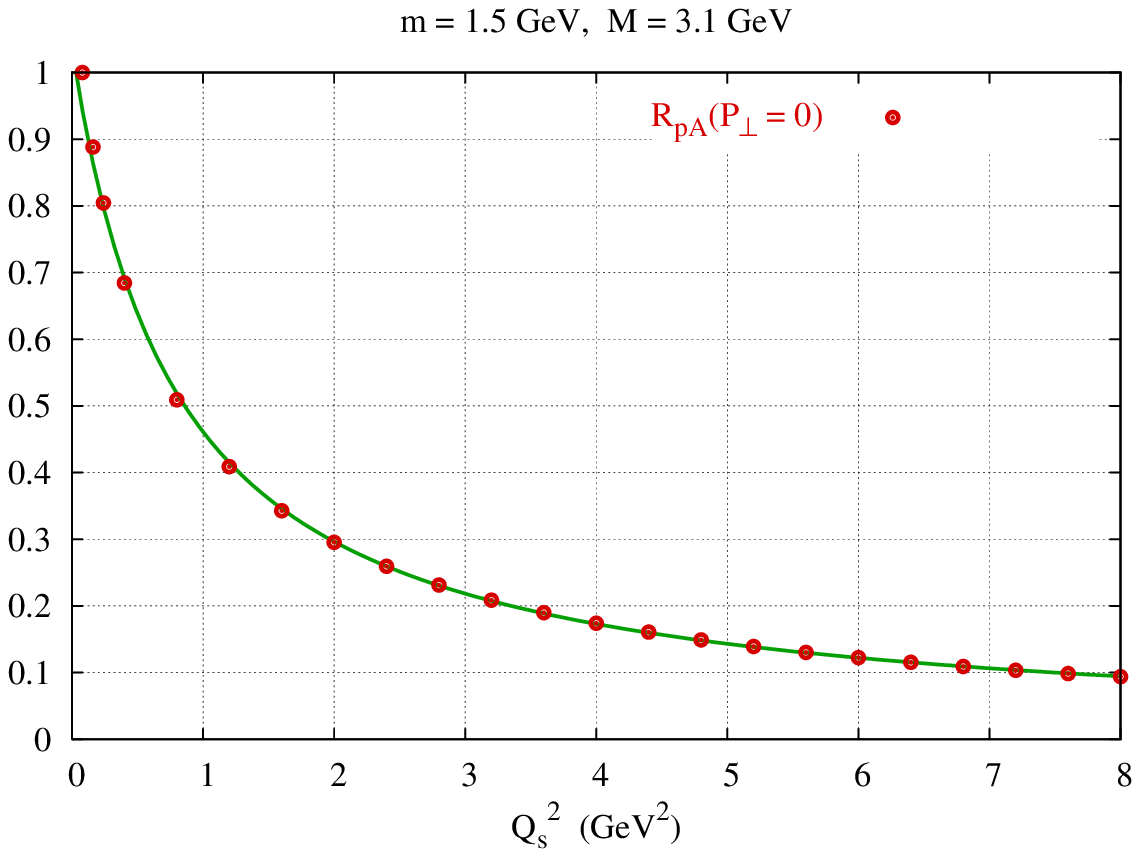}}
 \end{center}
 \caption{\label{fig:pair-Rpa} Left: nuclear modification factor for
 the production of $Q\overline{Q}$ pairs, as a function of their
 transverse momentum. Right: dependence of the modification factor at
 low $P_\perp$ as a function of $Q_s$. The solid line is fit by a
 function that behaves as $1/Q_s^2$.}
 \end{figure}
In the left plot of figure \ref{fig:pair-Rpa}, we display this ratio
as a function of $P_\perp$, for a given $Q_s$ and various fixed
invariant masses. One can see here a behavior which is very similar to
what was previously observed for gluon production in $pA$ collisions
(see \cite{BlaizGV1} for instance): there is a suppression at low
$P_\perp$ and an enhancement at high $P_\perp$, relative to perfect
scaling with the number of binary collisions.

The plot on the right of figure \ref{fig:pair-Rpa} shows how the value
of the suppression at small momentum changes with varying saturation
scale. We show this for an invariant mass just above the pair
production threshold because this is the dominant kinematical domain
contributing to the production of bound states (such as the $J/\psi$).
We observe that the pair yield, for large $Q_s^2$, behaves as
$1/Q_s^2$. Up to a logarithm (due to the small difference between
$Q_s^2$ and $\mu_{_A}^2$), this corresponds to a scaling like $1/L$
where $L$ is the longitudinal size of the nucleus.  This $1/L$
dependence is well known from the {\it
super-penetration}~\cite{Nemen1,LyuboP1} of electron-positron pairs
through metallic foils (often called the ``Chudakov effect") and was
noted previously in the QCD
case~\cite{Baym2,HeiseBBFS1,Fujii1,FujiiM1}, by looking at the
propagation of $Q\overline{Q}$ states through a random distribution of
color fields. This result is in contrast to the Glauber like form
$\exp(- C \rho \sigma L)$ that is often assumed, where $\sigma$ is the
inelastic $J/\psi$-nucleon cross-section, $\rho$ is the nuclear
density, and $C$ is constant number. Such an exponential form assumes
successive {\bf independent} collisions, and the decreasing
exponential is thus the probability that the $J/\psi$ has survived
after propagating through a length $L$ of nuclear matter. However, for
the values of $Q_s$ probed at present energies it may be difficult to
distinguish between the linear and exponential forms.
 \begin{figure}[htbp]
 \begin{center}
 \resizebox*{8cm}{!}{\includegraphics{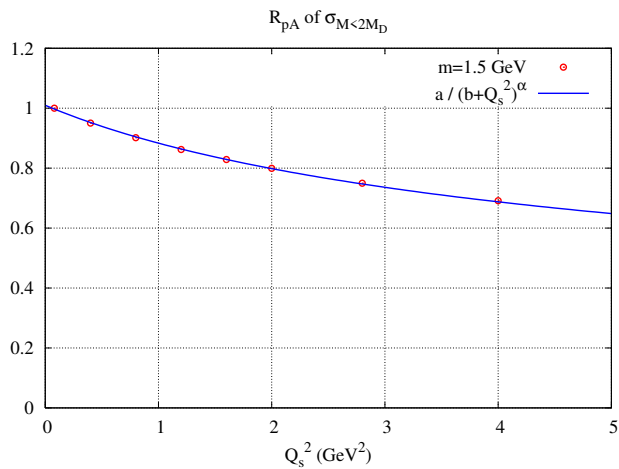}}
 \end{center}
 \caption{\label{fig:onium_cs} Ratio of the pair cross-sections with
 the invariant mass $M<2M_D$ for ``pp'' and ``pA.''  On the basis of
 the Color Evaporation Model, this is interpreted as the normal
 nuclear suppression factor of the charmonium. The parameters of the
 fitting curve are $a=1.51(6)~\mbox{GeV}^2$, $b=2.65(16)~\mbox{GeV}^2$ and
 $\alpha=0.417(15)$.}
 \end{figure}

To observe how this pattern of suppression for producing
$Q\overline{Q}$ pairs translates into the nuclear $J/\psi$
suppression, we will consider the phenomenological ``Color
Evaporation'' model (CEM)\footnote{See \cite{BrambA1} and references
therein for a review on quarkonium production.}. This model assumes
that the yield of $J/\psi$'s is obtained by integrating the yield of
$Q\overline{Q}$ pairs over the invariant mass of the pairs, from
production threshold of a charm quark pair to the threshold for the
production of a pair of $D$ mesons, up to an overall constant
prefactor. This can be expressed as
\begin{equation}
\frac{dN_{J/\psi}}{dY d^2\P_\perp}\empile{=}\above{CEM} F_{J/\psi}
\int_{4m^2}^{4m_{_D}^2}dM^2\;
 \frac{dN}{dY d^2\P_\perp dM^2}\; .
\end{equation}
If hadronization takes place outside of the nucleus (a legitimate
assumption for $J/\psi$ production at high energy), one can
assume that the prefactor $F_{J/\psi}$ is the same for $pA$ and $pp$
collisions. It therefore cancels out in the ratio $R_{pA}$. In figure
\ref{fig:onium_cs}, we represent the ratio $R_{pA}$ for $J/\psi$
production in the CEM~\footnote{The yield has been integrated over all
$\P_\perp$ before taking the ratio}, as a function of the saturation
momentum $Q_s$. The solid line represents a fit of this ratio by a
power of $Q_s$. One sees that the nuclear modification ratio
is well reproduced by a power law, $Q_s^{-0.8}$, namely $A^{-0.13}$ or
$1/L^{0.4}$. We get a smaller power of $Q_s$ relative to figure \ref{fig:pair-Rpa} simply because 
we here integrate over $P_\perp$.  The greater suppression at $P_\perp=0$
is attenuated by the enhancement present at higher $P_\perp$'s. 
This pattern of $J/\psi$ suppression is not very sensitive to the
precise mass of the $c$ quark.

\subsubsection{Dependence on the quark mass}
\begin{figure}[htbp]
\begin{center}
\resizebox*{8cm}{!}{\includegraphics{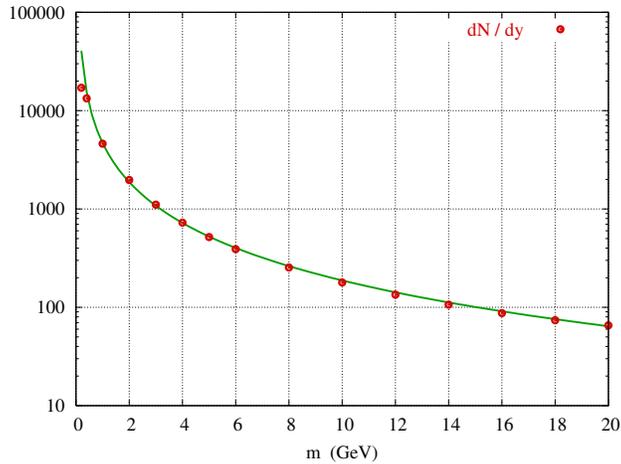}}
\end{center}
\caption{\label{fig:mass-dep} Dots: pair yield as a function of the
mass of the quarks. Solid line: fit by a function that behaves like
$\ln(m)/m^2$.}
\end{figure}
Finally, we shall discuss in the MV model, the dependence of the total
$Q\overline{Q}$ yield as a function of the mass of the quarks. This is
plotted in figure \ref{fig:mass-dep} (per unit rapidity; as
emphasized, there is no $Y$ dependence in the MV model).  One finds
that, for quarks masses larger than the saturation scale, the yield
behaves as $\ln(m)/m^2$, in agreement with what one expects from
perturbation theory. The leading $1/m^2$ simply comes from integrating
the $1/P_\perp^4$ behavior from $p_\perp\sim m$ to $\infty$. The
additional factor of $\ln(m)$ arises from the intermediate $P_\perp$
region, $Q_s \lesssim P_\perp \lesssim m$ where there is a deviation
from the $1/P_\perp^4$ behavior for masses below the saturation scale.
The yield therefore flattens out and does not grow as fast as $1/m^2$
anymore. This implies that the limit of zero quark mass is less
singular than it is at leading twist. In the latter case, the yield
would keep growing as $1/m^2$ all the way up into to the infrared
region.

 \subsection{Multiple scattering {\it a la} MV + quantum evolution {\it a la} BK}
\label{subsec:MVBK}

\subsubsection{Kinematics for small $x$ evolution and initial conditions}
 Our previous results, in the MV model, only included multiple scattering effects on
 quark pair production in proton-nucleus collisions. Quantum evolution effects
 due to gluon bremsstrahlung are not included. At high
 energies, gluons with a very small longitudinal momentum fraction $x$
 may be probed in the projectiles. It therefore becomes necessary to resum
 powers of $\alpha_s \ln(x_0/x)$. These corrections cause leading
 twist shadowing of gluon distributions. Because the initial condition
 for small $x$ quantum evolution is not the same in protons and 
 nuclei, quantum evolution is yet another effect -- in addition to
 the multiple scatterings already included in the MV model -- that
 makes the cross-sections in $pp$ and $pA$ collisions differ.

 As discussed in section 3, we will study these quantum evolution
 effects in the framework of the Balitsky-Kovchegov (BK) equation. We
 noted there that the nuclear saturation scale, obtained from
 solutions of the BK equation, now depends on both $A$ and the
 rapidity of the gluons probed in the nucleus. The former dependence
 comes from the initial condition, which is taken to be the
 McLerran-Venugopalan model, while the energy dependence follows from
 solutions of the BK equation. With the approximations for the
 correlators obtained by substituting the BK result for the l.h.s. of
 eq.~(\ref{eq:BKapprox}) in eqs.~(\ref{eq:largeN}-\ref{eq:2C3C}), we
 are in a position to study how our previous multiple scattering
 results are modified by small $x$ quantum evolution.

In our kinematics, the $+$ component of the gluon  that comes from the
 proton is related to the $+$ momentum of the pair by the relation
\begin{equation}
k_1^+ = p^+ + q^+\; ,
\end{equation} 
and similarly the $-$ momentum taken from the nucleus is given by the
 relation
\begin{equation}
k_2^- = p^- + q^-\; . 
\end{equation}
The longitudinal momentum fractions of the gluons probed in the proton
and in the nucleus -- respectively $x_p$ and $x_{_A}$ -- are related
to the $+$ component of the momentum of the proton and to the $-$
component of the momentum of a nucleon in the nucleus by~:
\begin{equation}
x_p = \frac{k_1^+}{P_p^+}\quad,\qquad
x_{_A} = \frac{k_2^-}{P_{_A}^-}\; .
\end{equation}
Using $P_p^+=P_{_A}^-=\sqrt{s/2}$, we can express these momentum
fractions in terms of the transverse momentum of the quark and the
antiquark and their rapidities (see the discussion following
eq.~(\ref{eq:pair-kinematics})) as 
 \begin{eqnarray}
 x_{p,_A} = \frac
{{\omega_p\, e^{\pm y_p} + \omega_q\, e^{\pm y_q} }}{ \sqrt{s} }\; ,
 \label{eq:x12}
 \end{eqnarray}
where $\omega_p^2\equiv \p_\perp^2+m^2$ and $\omega_q^2\equiv
\q_\perp^2+m^2$. Alternatively, we can also express the momentum fractions in terms of
the kinematical parameters of the pair, $P_\perp,M,Y$ as
\begin{equation}
x_{p,_A}=\sqrt{\frac{P_\perp^2+M^2}{s}}\;e^{\pm Y}\; .
\label{eq:x12-1}
\end{equation}
This latter form of $x_{p,A}$ makes obvious the fact that when we
integrate over $\tilde{q}$ and $\phi$ in
eq.~(\ref{eq:pair-cross-section}), the values of $x_p$ and $x_{_A}$
remain constant. This is very similar to gluon
production, where the values of $x_{p,_A}$ are also fully determined by the
final state\footnote{The analogy between the two is
not accidental and follows from the fact that all the
produced particles are measured in both instances. In contrast, in single quark production, one integrates out the
phase-space of the antiquark thereby allowing the values of $x_{p,_A}$ to
vary.}.

 Our approach has an important caveat that we should discuss before
 presenting the results. We have argued elsewhere that the MV model, 
 used here as the initial condition when solving the BK equation, is a
 good description of a nucleus at moderate value of $x$. This value is often estimated~\footnote{This arbitrariness is similar 
 to that of the initial condition $Q_0^2$ in DGLAP evolution. In both instances, asymptotic results are independent of 
 the initial condition.} to be of order
 of $x_0=10^{-2}$. By solving the BK equation, one obtains the value
 of the gluon distributions at any value of $x$ smaller than this
 $x_0$. However, in the calculation of the $Q\overline{Q}$
 cross-section, depending on the kinematics of the final state, we may need the unintegrated gluon
 distributions for values of $x_p$ or $x_{_A}$ (or both) that are
 larger than $x_0$. From eq.~(\ref{eq:x12-1}), we observe that this will
 happen in one of the projectiles if $|Y|$ becomes large, or in both
 of them at moderate $Y$ and large $P_t$ or $M$. Therefore, the MV
 initial condition at $x=x_0$ supplemented by  BK evolution, is not
 sufficient to cover the entire kinematical range. One possibility would
 be to simply dismiss the points $P_t,Y,M$, where this problem occurs,
 on the grounds that these points are ``contaminated'' by large-$x$
 physics. Ideally, one should try to constrain the necessary gluon
 distributions at large $x$ from existing data. However, for the purposes of the
 qualitative discussion presented here, we
 follow an intermediate path: we make a simple ansatz for the gluon
 distributions that continues the MV initial condition at $x\ge x_0$.

Because the numerical evaluation of the pair cross-section is done with
eq.~(\ref{eq:cross-section-1}), we need only to make such an ansatz
for the proton unintegrated gluon distribution $\varphi_p$ and for the
3-point function $\phi_{_A}^{g,q\bar{q}}(\k_{2\perp};\k_\perp|x)$ in the nucleus. One generally
requires, from quark counting rules,  that for values of $x$ close to unity, all gluon
distributions vanish as $(1-x)^4$. We adopt here a simple
ansatz that implements this behavior; 
\begin{equation}
\mbox{for\ \ }x\ge x_0\;,\qquad
\phi_{_A}^{g,q\bar{q}}(\k_{2\perp};\k_\perp|x)
\equiv
\left(\frac{1-x}{1-x_0}\right)^4\;
\phi_{_A}^{g,q\bar{q}}(\k_{2\perp};\k_\perp|x_0)\; ,
\label{eq:large-x-extra}
\end{equation}
where the denominator $(1-x_0)^4$ ensures the continuity of
$\phi_{_A}^{g,q\bar{q}}(\k_{2\perp};\k_\perp|x)$ at $x=x_0$. We 
make the same ansatz for $\varphi_p$.  Because the details of how these gluon distributions vanish when 
$x\to 1$ are somewhat arbitrary, readers should note that results for the
cross-section in kinematic regions dominated by large values of $x$ (in one or both of the
projectiles) contain a large systematic uncertainty. 

This uncertainty is reduced when the center of mass
energy $\sqrt{s}$ of the collision increases because the values of $x_{p,_A}$ decrease and the 
corresponding physics is less sensitive to the initial conditions. To ensure a significant region in the kinematical
 parameters $P_\perp,Y,M$ of the pair where the cross-section is not affected by this
arbitrariness, all  numerical calculations presented in the
following subsections have been performed for the center of mass energy of
the LHC, $\sqrt{s}=8.5~$TeV. At RHIC energies, {\it if $x_0 = 10^{-2}$}, there is no region in 
phase-space which is not contaminated by the large $x$ behavior of
one (or both) of the projectiles. Therefore, a realistic
description of RHIC results in this framework would require a less simplistic ansatz than 
eq.~(\ref{eq:large-x-extra}). Alternately, it may very well be the case that MV initial conditions may be applicable 
in large nuclei at larger values of $x_0$. A systematic study is required but is beyond the scope of this paper.

\subsubsection{Rapidity dependence} 
\begin{figure}[htbp]
 \begin{center}
  \resizebox*{8cm}{!}{\rotatebox{-90}{\includegraphics{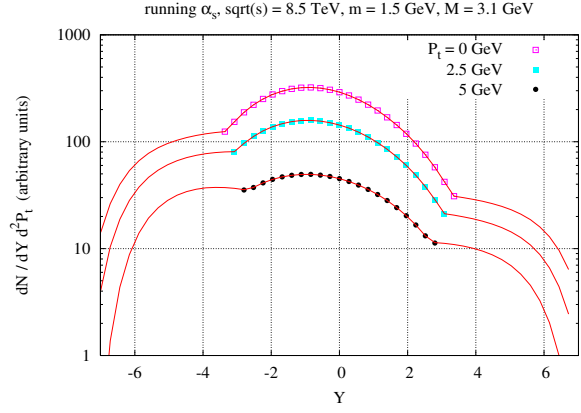}}}
 \end{center}
 \caption{\label{fig:rapidity} The rapidity distribution of
   quark--anti-quark pairs in pA collisions, for an invariant mass
   $M=3.1~$GeV and various values of the transverse momentum
   $P_\perp$. The dots indicate the region in $Y$ which is not
   sensitive to the extrapolation done at large $x$. (See
   eq.~(\ref{eq:large-x-extra})).}
 \end{figure}

We shall now use the solution of the BK equation with the
 extrapolation done in eq.~(\ref{eq:large-x-extra}),  to
 compute the rapidity dependence of the pair cross-section in eq.~(\ref{eq:pair-cross-section}).
 In fig.~\ref{fig:rapidity}, we show the rapidity distribution of
 $c{\bar c}$ pairs, for a fixed value of the invariant mass
 $M=3.1$~GeV, and several values of the transverse momentum of the
 pair. The fact that the first derivative in $x$ of
 the unintegrated gluon distributions is not preserved by the ansatz~\footnote{We have also tried an alternative 
 large $x$ ansatz of the form $A x^\alpha (1-x)^4$, where $A$ and $\alpha$ are constrained to ensure that the 
 unintegrated gluon distribution and its first derivative are continuous at $x=x_0$. This ansatz in turn leads to 
 odd behavior of the unintegrated distributions at large $p_\perp$.} of 
 eq.~(\ref{eq:large-x-extra}) shows up in the cross-section as
 a discontinuity in the slope in $Y$ of $dN/dYdM^2 d^2\P_\perp$. On
 the plots, we have indicated the region in $Y$ affected by this
 extrapolation by plotting only a solid line in these regions (their
 boundaries can be easily determined from eq.~(\ref{eq:x12-1})). The
 central region in rapidity - unaffected by this extrapolation -
 shrinks as one increases the transverse momentum of the pair, or its
 invariant mass.

\subsubsection{Cronin effect and shadowing}
\begin{figure}[htbp]
 \begin{center}
 \hfill
 \resizebox*{5.5cm}{!}{\rotatebox{-90}{\includegraphics{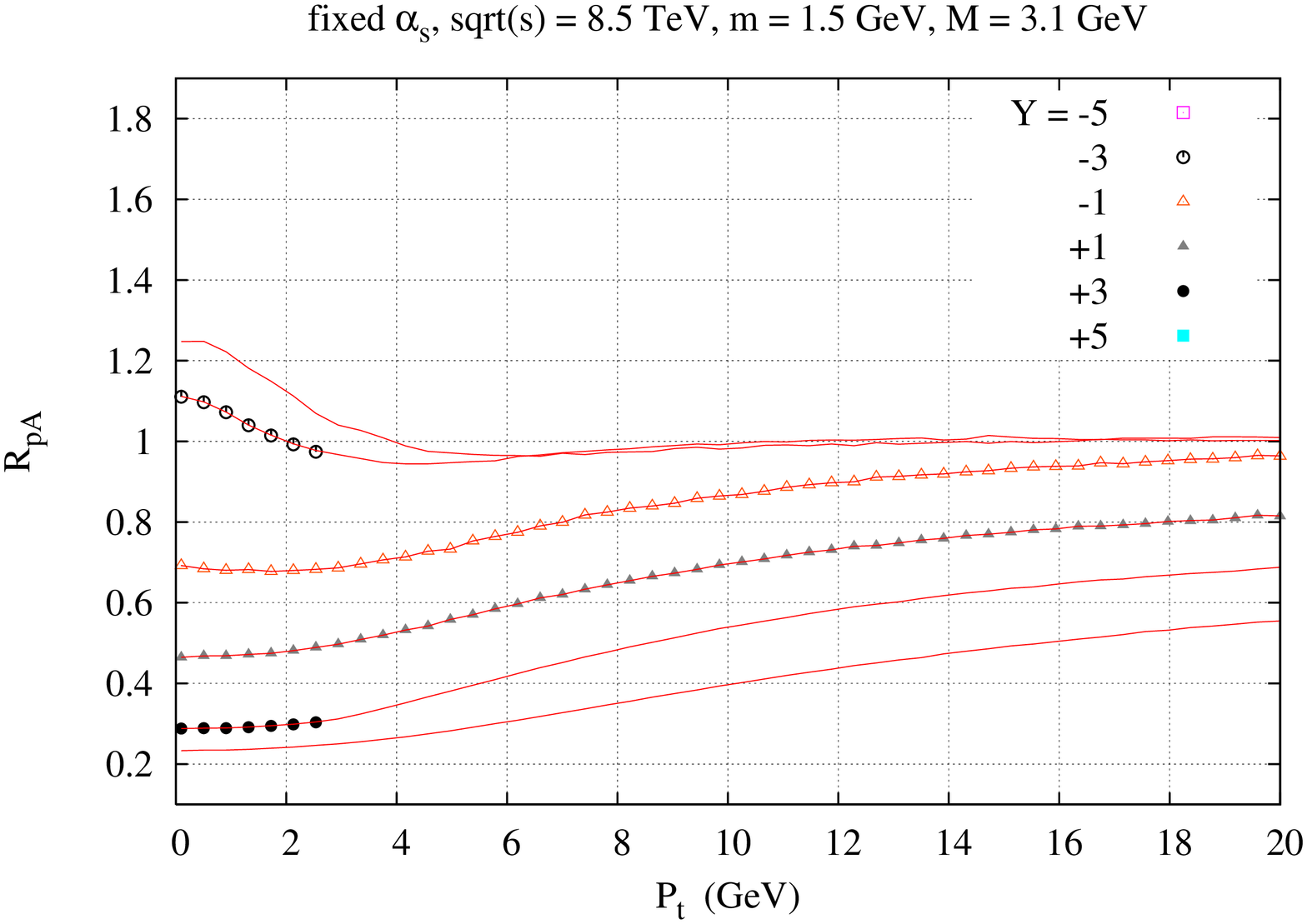}}}
 \hfill
 \resizebox*{5.5cm}{!}{\rotatebox{-90}{\includegraphics{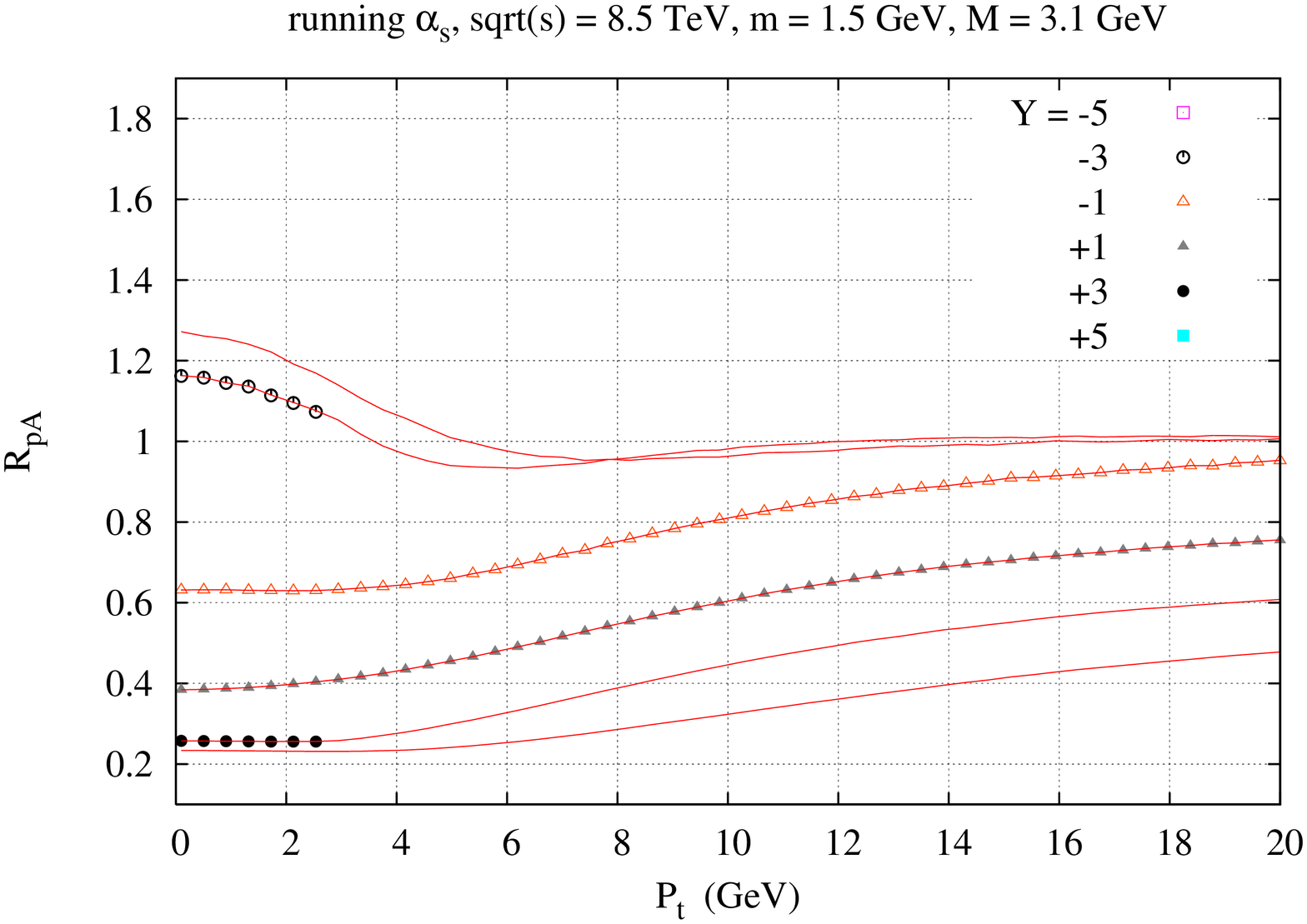}}}
 \hfill
 \end{center}
 \caption{\label{fig:rpa-BK} Nuclear modification factor for
 $Q\overline{Q}$ pairs at LHC center of mass energy, as a function of
 the transverse momentum of the pair. The pair invariant mass is 
 just above the charm pair threshold ($M=3.1$~GeV) and the rapidity is varied from
 $-5$ to $+5$. Left: the $x$ dependence is obtained from the BK
 equation with fixed coupling ($\alpha_s N_c/\pi=0.2$). Right: the $x$
 dependence is obtained from the BK equation with running
 coupling. Regions on the plot represented by only a solid line and no dots are
 sensitive to the large-$x$ extrapolation done in
 eq.~(\ref{eq:large-x-extra}).}
 \end{figure}

To study deviations from the scaling with the number of
binary collisions, one can compute the ratio $R_{pA}$ defined in
eq.~(\ref{eq:rpa-def}). In figure \ref{fig:rpa-BK}, we display this
ratio as a function of $P_\perp$ for different values of the rapidity
(again the invariant mass is set to $M=3.1$~GeV). On the left, the
cross-section has been computed with the solution of the BK equation
with fixed coupling, while on the right the BK equation was solved
with a running coupling. There are small
quantitative differences between fixed and running coupling, but the
overall qualitative behavior of $R_{pA}$ is unaffected.
As one can see, the only region for which the ratio $R_{pA}$ goes over
1 is at the most negative rapidities. As the rapidity increases above
$Y=-3$, $R_{pA}$ decreases significantly, and remains below 1 at all
$P_\perp$'s.
\begin{figure}[htbp]
 \begin{center}
  \resizebox*{8cm}{!}{\rotatebox{-90}{\includegraphics{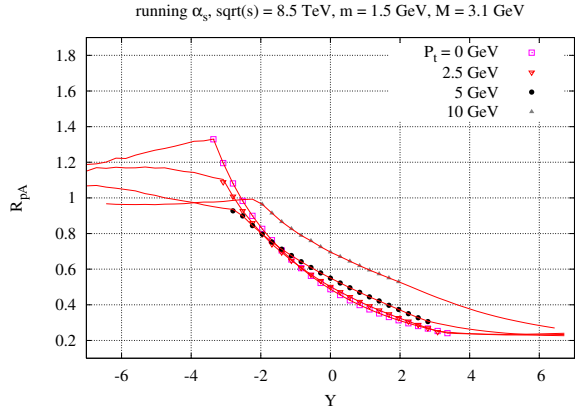}}}
 \end{center}
 \caption{\label{fig:rpa-Y} The rapidity dependence of the nuclear
 modification ration, at fixed $M$ and for several values of
 $P_\perp$. The regions affected by the large $x$ extrapolation are
 indicated by the absence of dots.}
 \end{figure}
An alternative way to view the same information is to display $R_{pA}$ as a
function of the rapidity $Y$, for fixed values of $P_\perp$, as in
figure \ref{fig:rpa-Y}. (Note that for a limited region in $Y$, and for fixed $P_\perp$, it is proportional to the widely used variable $x_F$.) 
On this plot, one sees a rapid decrease of
$R_{pA}$ with $Y$. In our computation, this drop seems to start around
$Y=-3$.  Keep in mind however that the lower rapidity region is 
plagued by artifacts introduced by the large-$x$ extrapolation--the drop 
may in fact start earlier. Moreover, one sees on this plot
that the decrease of $R_{pA}$ with $Y$ is very rapid at negative $Y$,
and slows down at positive rapidities. This is reminiscent of the very
fast disappearance of the Cronin peak in the study of $R_{pA}$ for
gluons via the BK equation \cite{AlbacAKSW1,KharzKT1}. One also
observes that the pattern of this suppression is almost independent of
$P_\perp$ at small $P_\perp$, and that the $P_\perp$ dependence
becomes manifest only above $P_\perp\sim 5~$GeV. These results are qualitatively similar to 
those observed for the rapidity dependence of the nuclear modification factor at lower energies.

 \section{Conclusions}
In this paper, we have explored the formalism developed in
ref.~\cite{BlaizGV2} for pair production in proton nucleus collisions.
We presented explicit results for two models, each of which
corresponds to a particular limit of high parton densities realized in
the Color Glass Condensate formalism of high energy QCD. In the first
case, the McLerran-Venugopalan model, multiple scattering effects on
pair production were studied in the absence of shadowing effects. We
obtained results in this model for the pair cross-sections as a
function of the invariant mass, the transverse momenta of the pairs,
as well as their dependence on nuclear size and the quark mass.  For
large transverse momenta or invariant masses, both the power law
dependence on these scales, and the accompanying logarithms can be
interpreted in terms of the leading kinematic behavior in the
collinear factorization framework of perturbative QCD. We also
studied, in the MV model, the extent of the violation of $k_\perp$
factorization.

We next considered a model, where both multiple scattering and small
$x$ quantum evolution (shadowing) effects are included. In the limit
of large $N_c$ and large $A$, the energy evolution of the multi-parton
correlators in pair cross-sections can be described in terms of the
Balitsky-Kovchegov equation. We solved the BK equation numerically,
for both fixed and running coupling, in order to determine the
rapidity distribution of pairs as well as the evolution of the nuclear
modification factor $R_{pA}$ as a function of rapidity. As in the
gluon case, for fixed $P_\perp$, one notes a rapid initial depletion
of $R_{pA}$ with rapidity followed by a much slower depletion at
larger rapidities. We emphasize that at RHIC energies our results may
be sensitive to particular ans{\"a}tze for unintegrated gluon
distributions at large $x$; at LHC energies, this sensitivity is
considerably weaker.

The studies in this paper provide insight into the systematics of pair
production and the dependence of the results on the quark masses, the
system size and the center of mass energy. Detailed phenomenological
studies of $J/\psi$ and open charm production are underway and will be
reported on in a followup to this work~\cite{FujiiGV2}.

While this manuscript was being prepared for publication,
ref.~\cite{KovchT2} appeared. The authors of ref.~\cite{KovchT2}
correctly argue that the results of ref.~\cite{BlaizGV2} can be
extended to include small $x$ quantum evolution effects. This is
precisely what is done here; quantum evolution effects are included by
solving the BK equation for the unintegrated gluon distributions in
eq.~\ref{eq:corr-nucleus}. Quantitative results from the solution of
the BK equation for the evolution of pair cross-sections with
energy/rapidity are presented in section~\ref{subsec:MVBK}.

\section*{Acknowledgements}
HF and FG would like to thank the RIKEN-BNL center and BNL Nuclear
Theory respectively for their hospitality. Likewise, RV would like to
thank Service de Physique Th\'eorique, Saclay. RV's research was
supported by DOE Contract No. DE-AC02-98CH10886. HF's work was
supported by the Grants-in-Aid for Scientific Research (\# 16740132).


\end{document}